\documentclass[10pt,a4paper]{article}

\usepackage{jheppub}

\usepackage{color}
\usepackage{colordvi}
\usepackage{changes}

\usepackage{epsfig,epsf}
\usepackage{amsmath}
\usepackage{amsthm}
\usepackage{amsfonts}
\usepackage{amssymb}
\usepackage{dsfont}
\usepackage{epstopdf}
\usepackage{multirow}
\usepackage{braket}
\usepackage{tabularx}

\usepackage{rotating}

\usepackage{marvosym}

\usepackage{todonotes}

\usepackage{subcaption}
\usepackage{array}   
\newcolumntype{C}{>{$}c<{$}}

\usepackage{slashed}

\usepackage[active]{srcltx}

\newcommand{\p}{\partial}
\newcommand{\s}[1]{\slashed{#1}}

\newcommand{\w}[1]{\widetilde{#1}}
\newcommand{\f}[1]{\mathcal{#1}}

\newcommand{\sbar}[1]{\slashed{\bar{#1}}}

\newcounter{RSQ}

\usepackage{soul}

\newcommand \intR {\int_{-\infty}^{\infty}}

\def\II{\hbox{{1}\kern-.25em\hbox{l}}}

\def\II{\hbox{{1}\kern-.25em\hbox{l}}}

\title{
All order factorization for virtual Compton scattering at next-to-leading power}

\author[a]{Jakob Schoenleber,}
\author[b]{Robert Szafron}
\affiliation[a]{
   RIKEN BNL Research Center, Brookhaven National Laboratory, Upton, NY 11973, USA}
\affiliation[b]{
   High Energy Theory Group, Physics Department, Brookhaven National Laboratory, Upton, NY 11973, USA}

\emailAdd{jakob.schoenleber@gmail.com}
\emailAdd{rszafron@bnl.gov}

\abstract{We discuss all-order factorization for the virtual Compton process at next-to-leading power (NLP) in the $\Lambda_{\rm QCD}/Q$ and $\sqrt{-t}/Q$ expansion (twist-3), both in the double-deeply-virtual case and the single-deeply-virtual case. We use the soft-collinear effective theory (SCET) as the main theoretical tool. We conclude that collinear factorization holds in the double-deeply virtual case, where both photons are far off-shell.
The agreement is found with the known results for the hard matching coefficients at leading order $\alpha_s^0$, and we can therefore connect the traditional approach with SCET. In the single-deeply-virtual case, commonly called deeply virtual Compton scattering (DVCS), the contribution of non-target collinear regions complicates the factorization. These include momentum modes collinear to the real photon and (ultra)soft interactions between the photon-collinear and target-collinear modes. However, such contributions appear only for the transversely polarized virtual photon at the NLP accuracy and in fact it is the only NLP $\sim (\Lambda_{\rm QCD}/Q)^1 \sim (\sqrt{-t}/Q)^1$ contribution in that case. 
We therefore conclude that the DVCS amplitude for a longitudinally polarized virtual photon, where the leading power $\sim (\Lambda_{\rm QCD}/Q)^0 \sim (\sqrt{-t}/Q)^0$ contribution vanishes, is free of non-target collinear contributions and the collinear factorization in terms of twist-3 GPDs holds in that case as well.  
}


\begin{document}
\maketitle

\section{Introduction}

The GPDs ~\cite{Muller:1994ses, Ji:1996ek, Ji:1996nm, Radyushkin:1997ki} have a rich physical interpretation in terms of transverse spatial probability distribution of partons with a given longitudinal momentum fraction~\cite{Burkardt:2002hr} and are essential for studying the decomposition of the proton spin~\cite{Ji:1996ek} and various inter- and multi-parton correlations inside the hadrons. GPD studies are subject to active research in nuclear and high-energy physics and have been stated as a significant science goal for the planned Electron-Ion collider at Brookhaven National Laboratory~\cite{AbdulKhalek:2021gbh, AbdulKhalek:2022hcn}.

The deeply virtual Compton scattering (DVCS)~\cite{Ji:1996ek, Ji:1996nm} is considered the ``golden'' channel for experimentally accessing GPDs. This process is defined as electroproduction of a real photon from the nucleon
\begin{figure}
\centering
\includegraphics[scale=.3]{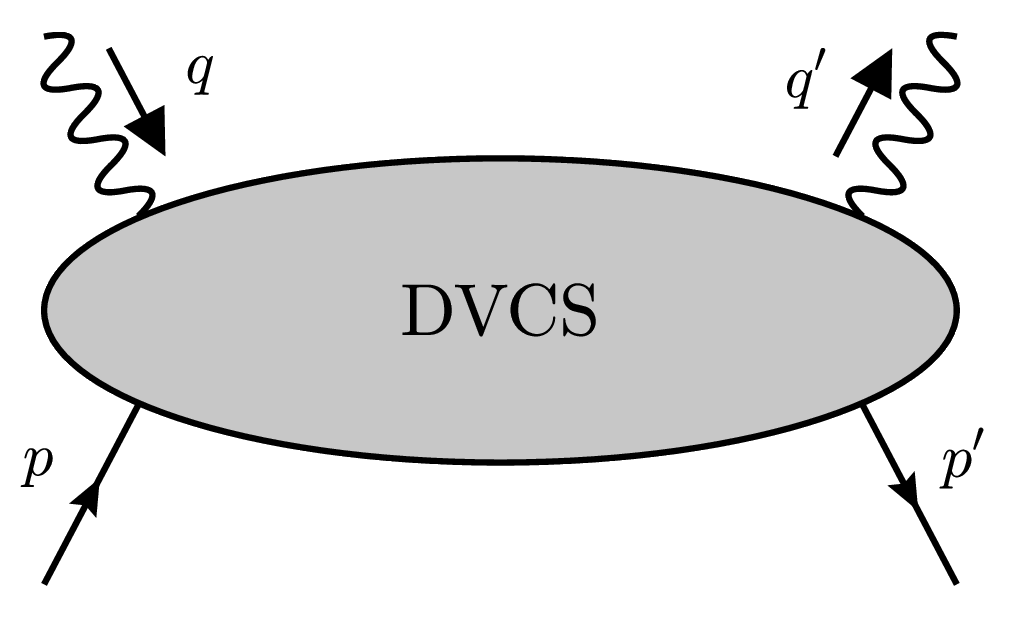}
\caption{The DVCS process.}
\label{fig: DVCS}
\end{figure}
\begin{align}
\gamma^*(q) + N(p) \rightarrow \gamma(q') + N'(p'),
\end{align}
where $\gamma^*$ is the highly virtual photon exchanged between a lepton and the nucleon. Its virtuality $-q^2 > 0$ provides the hard scale of the processes, and it is assumed to be much larger than the scale of strong interactions $\Lambda_{\rm QCD}$. This strong scale separation allows for factorization of the short-distance hard physics contained in the coefficient functions and long-distance non-perturbative effects encoded in GPDs. 
The QCD input needed to describe DVCS is entirely encoded in the hadronic tensor
\begin{align}
T^{\mu \nu} = i \int d^4z\, e^{-iq\cdot z} \bra{p'} T\{ j^{\mu}(z) j^{\nu}(0) \} \ket p, 
\end{align}
where $j^{\mu} = \sum_q e_q \bar \psi_q \gamma^{\mu} \psi_q$ is the electromagnetic current. 
Factorization and perturbative behavior of DVCS has been comprehensively understood at leading power: All-order factorization at $(\Lambda_{\rm  QCD}/Q)^0$ has been proven in \cite{Radyushkin:1997ki, Collins:1998be, Ji:1998xh, Bauer:2002nz} and the coefficients function are known to the next-to-next-to-leading order $\alpha_s^2$~\cite{Braun:2022bpn, Ji:2023xzk}. A great deal is known beyond leading power due to the operator product expansion; for example,~\cite{Braun:2012hq} and the recent development \cite{Braun:2022qly}. 

In this work, we focus on next-to-leading power (NLP) effects\footnote{Compton scattering at NLP has been studied extensively in the late 90's and early 00's. For an detailed account of the historical development we refer to \cite{Diehl:2003ny}, chapter 6.}, which are suppressed by $(\Lambda_{\rm QCD}/Q)^1$ compared to the leading power. At this accuracy, the DVCS amplitude has been hypothesized to factorize into twist-3 GPDs, which have a rich physical interpretation in themselves, see, e.g.,~\cite{Hatta:2012cs, Hatta:2024otc}. 
The corresponding leading  $\alpha_s^0/Q$ coefficient functions have been known for a long time and even a part of the next-to-leading order $\alpha_s^1/Q$ contribution has been calculated in~\cite{Kivel:2003jt}.  For a more recent discussion, see~\cite{Aslan:2018zzk}.

However, the all-order factorization of DVCS at NLP has never been discussed in the same level of detail as that of the leading power. The important subtleties that cast doubts about the factorization in its hypothesized form are due to non-target-collinear regions that arise when the outgoing photon becomes real, therefore introducing a second collinear direction. 
In particular, if the regions where partons become soft contribute to the power accuracy in question, one may expect endpoint-like divergences of the convolution integral of the hard coefficient functions with the non-perturbative objects. This usually implies that the IR singularities are not adequately factored into these non-perturbative objects. While this does not necessarily mean that factorization in the broad sense is broken (this can happen for example if Glauber regions are present), it might mean that there are additional contributions that require more sophisticated techniques (see~\cite{Liu:2019oav,Beneke:2020ibj,Hurth:2023paz} and~\cite{Beneke:2022obx,Liu:2020wbn,Liu:2020tzd,Liu:2022ajh,Beneke:2019kgv,Cornella:2022ubo} for related discussion). The observed absence of endpoint-like divergences in DVCS at NLP at tree-level \cite{Kivel:2000cn} (and partly at one-loop~\cite{Kivel:2003jt}) does indicate that factorization holds in its assumed form. Still, it provides by no means a level of confidence associated with established factorization analyses. Indeed, it is not clear whether regions where partons become (ultra)soft necessarily leads to endpoint-like divergences, even if they contribute to the power accuracy in question. On the other hand, if the (ultra)soft endpoint region is suppressed relative to the leading collinear region there can not be an endpoint divergence. We will show that this is the case for a longitudinally polarized photon, which implies the absence of endpoint divergences at NLP and to all orders in that case.

The soft-collinear effective theory (SCET)~\cite{ Bauer:2000yr, Bauer:2001ct, Bauer:2001yt, Bauer:2002nz} enables a systematic study of power corrections without appealing to the operator product expansion. In this work, we use the position space formulation~\cite{Beneke:2002ph, Beneke:2002ni} and we will restrict ourselves to the infrared (IR) modes\footnote{We consider two light-cone vectors
\begin{align}
n^{\mu} = (1,0,0,1)^{\mu}, \qquad \bar n^{\mu} = (1,0,0,-1)^{\mu}.
\end{align}
An arbitrary vector $v^{\mu}$ can be decomposed as
\begin{align}
v^{\mu} = \bar n \cdot v \, \frac{n^{\mu}}{2} + n \cdot v \, \frac{\bar n^{\mu}}{2} + v_{\perp}^{\mu}.
\end{align}
Frequently, we will denote four-vectors by their light-cone components as
\begin{align}
v = (n \cdot v, \bar n \cdot v, v_{\perp}).
\end{align}
We commonly use the definitions
\begin{align}
g_{\perp}^{\mu \nu} &= g^{\mu \nu} - \frac{n^{\mu} \bar n^{\nu} + n^{\nu} \bar n^{\mu}}{2}, \qquad\varepsilon_{\perp}^{\mu \nu} = \varepsilon^{\mu \nu \mu' \nu'}  \frac{ n_{\mu'} \bar n_{\nu'} }{2}.
\end{align}} 
\begin{align}
\text{collinear}: \qquad &p_c \sim (\lambda^2, 1, \lambda)Q, \notag
\\
\text{anti-collinear}: \qquad &p_{\bar c} \sim (1, \lambda^2, \lambda)Q,
\label{eq: regions}
\\
\text{ultrasoft}: \qquad &p_{us} \sim (\lambda^2, \lambda^2, \lambda^2)Q. \notag
\end{align}
Here $\lambda$ is a dimensionless power counting parameter that equals some fixed momentum scale, in this case either $\Lambda_{\rm QCD},\,m  = \sqrt{p^2} = \sqrt{p'^2}$ or $\sqrt{-t} = \sqrt{-(p-p')^2}$, divided by $Q$. In other words, $\lambda \rightarrow 0$ parametrizes the limit $Q \rightarrow \infty$ with the different scales fixed. It is not necessary to further specify the definition of $\lambda$ unless one wants to investigate potentially large-scale ratios in certain kinematic limits such as $Q \gg \sqrt{-t} \gg \Lambda_{\rm QCD}$. This kinematic limit is also connected to the chiral and trace anomalies of QCD, which was subject to some recent work in a related context \cite{Tarasov:2020cwl, Tarasov:2021yll,Bhattacharya:2022xxw, Bhattacharya:2023wvy}. We will not discuss such issues in this work and assume that all low energy scales are parametrically equal.

We choose a frame such that the approximate direction of the in- and outgoing proton is along $n$ and the approximate direction of the outgoing photon is along $\bar n$. This means we will refer to the proton momentum as collinear and the outgoing photon as anti-collinear. 
In the modern effective field theory (EFT) approach, every momentum mode of a field that corresponds to a different scale is described by separate fields. The hard modes are integrated-out, and SCET contains only fields corresponding to collinear, anti-collinear, and ultrasoft modes:\footnote{More precisely, this decomposition is to be understood as in the method of regions framework \cite{Beneke:1997zp, Jantzen:2011nz}. Moreover, the precise decomposition in QCD, valid beyond leading power and preserving gauge symmetry, is more involved~\cite{Beneke:2002ni}.}
\begin{align}
\phi = \phi_c + \phi_{\bar c} + \phi_{us} \,.
\end{align}
The Lagrangian is subsequently expanded in $\lambda$, so each term has strictly homogeneous scaling in the expansion parameter. The resulting Lagrangian up to next-to-leading power and all corresponding definitions are collected in Appendix \ref{sec: SCETI L}.

Some comments are in order regarding the presence of the ultrasoft mode, instead of \textit{soft} mode
\begin{align}
\text{soft}: \qquad p_s \sim (\lambda, \lambda, \lambda) Q.
\end{align}
For exclusive processes the more ``physical'' low-energy mode is in some sense the soft mode. This is because the length scale at which the reconfining of a parton, which is hit by the hard probe, happens should not exceed the size of the hadron $\sim \Lambda_{\rm QCD}^{-1} \sim (\lambda Q)^{-1}$. Ultrasoft radiation propagates over length scales $\lambda^{-2}Q$ which is much larger than the size of the hadron. In order words we expect that confinement effectively ``cuts off'' virtualities at $\sim \Lambda_{\rm QCD}^2$ for exclusive processes, making ultrasoft modes unphysical in this sense. However, for the sake of simplicity, we will still use ultrasoft as the designated low-energy mode for this work, noting that this does not change the main conclusion of this work. In particular, the matching coefficients in SCET do not depend on the detailed arrangement of the low energy modes.


Another possible momentum space region that should be mentioned is the Glauber region, where $\bar n \cdot p_G\, n \cdot p_G \ll p_{G\perp}^2$. This region is notoriously dangerous for factorization and their implementation in SCET, while well understood~\cite{Rothstein:2016bsq}, is complicated. However, it is a well-established~\cite{Collins:2011zzd} claim that a necessary condition for the Glauber pinch is the presence of at least two collinear external particles, which are in different collinear directions, in both the initial and final state (given that there are no soft external particles). Since the virtual Compton process can not meet this criterion, it is safe to assume that no Glauber pinch can occur. We note that in the context of GPD physics, Glauber regions have been identified in~\cite{Nabeebaccus:2023rzr, Braun:2002wu}. These processes satisfy the above necessary condition for the Glauber pinch.

So far, SCET has been applied to describe GPDs in a single work~\cite{Bauer:2002nz}, where only the leading power factorization is discussed. In this work, we extend the analysis to NLP. The arguments that require the most new development are needed to deal with the collinear and ultrasoft modes that superficially contribute at NLP. 

We start by analyzing a one-loop graph in section~\ref{sec: example} and show that, for this example, the anti-collinear and ultrasoft contributions are next-to-next-to-leading power (NNLP). In section~\ref{sec: factorization}, we extend the analysis to all orders using SCET. In the subsection~\ref{sec: nontarget regions}, we discuss the contributions from the anti-collinear and ultrasoft region at the operator level.

In section \ref{sec: calc}, we perform the tree-level matching, which is a straightforward exercise. Then, in section \ref{sec: old approach}, we show that the SCET formulation reproduces the established result \cite{Kivel:2000cn} in that it yields the same twist-3 GPDs with the same coefficient functions. In the literature, many different frames are used to describe DVCS. One can translate between the frames using reparametrization invariance (RPI) symmetry of SCET. We use the frame introduced in \cite{Braun:2012hq} throughout most of this work, but we also convert to the frames of \cite{Kivel:2000cn} and \cite{Belitsky:2005qn} by using RPI in section \ref{sec: RPI}.

Since it does not cost much effort, we simultaneously discuss the case where both photons are far off-shell, known as double-deeply virtual Compton scattering (DDVCS). However, we always have in mind the conceptually more riveting case of DVCS, where the outgoing photon is real.

\section{Kinematics and reference frame}
We use the following definitions
\begin{align}
P^{\mu} = \frac{p^{\mu}+p'^{\mu}}{2}, \qquad \Delta = p' - p, \qquad Q^2 = -\frac{1}{4} (q+q')^2, \qquad t = \Delta^2 \,,
\end{align}
and we define
\begin{align}
\rho &= - \frac{(q+q')^2}{4 P \cdot (q+q')}.
\end{align}
Note that $\rho$ reduces to $x_B = \frac{Q^2}{2p \cdot q}$ in the forward case $p = p'$. 

 The skewness $\eta$ is defined  as
\begin{align}
\eta = - \frac{\Delta \cdot \bar n}{2P \cdot \bar n}\,,
\end{align}
and is not a Lorentz-scalar since $\bar n^{\mu}$ is a fixed reference vector that does not transform under Lorentz transformations. Indeed, definitions of $\eta$ in the literature differ by $\f O(\lambda^2)$ terms. 
DVCS is obtained by taking $\rho = \eta$, as will be manifest from the explicit parametrization of the momenta below.

As mentioned, we always assume collinear scaling $p, p', \Delta, P \sim (\lambda^2, 1 , \lambda)Q$ and this implies that $1-\eta \sim \eta \sim 1$. Furthermore, we assume that $q' \sim (1, \lambda^2, \lambda)Q$ for DVCS. Various frames are used in the DVCS literature to satisfy these conditions. We consider the following three:
\begin{itemize}

\item[1.] The Compton frame $P_{\perp} = q_{\perp} + q_{\perp}' = 0$, which is used for example in \cite{Ji:1998xh} and \cite{Belitsky:2005qn}. We have
\begin{align}
\Delta_{\perp}^2 = 4m^2 \eta^2 + t(1- \eta^2)\,,
\end{align}
and
\begin{align} \notag
p^{\mu} &= (1+\eta) \bar n \cdot P \frac{n^{\mu}}{2} + (1-\eta) \frac{m^2 - t/4}{\bar n \cdot P} \frac{\bar n^{\mu}}{2} - \frac{1}{2} \Delta_{\perp}^{\mu},
\\ \notag
p'^{\mu} &= (1-\eta) \bar n \cdot P \frac{n^{\mu}}{2} + (1+\eta) \frac{m^2 - t/4}{\bar n \cdot P} \frac{\bar n^{\mu}}{2} + \frac{1}{2} \Delta_{\perp}^{\mu},
\\
q^{\mu} &= - (\rho + \eta) \bar n \cdot P \frac{n^{\mu}}{2} + \frac{Q^2}{\rho \bar n \cdot P} \frac{\bar n^{\mu}}{2} + \frac{1}{2} \Delta_{\perp}^{\mu} + \mathcal{O}\left(\frac{t}{Q},\frac{m^2}{Q}\right),
\\
q'^{\mu} &= - (\rho - \eta) \bar n \cdot P \frac{n^{\mu}}{2} + \frac{Q^2}{\rho \bar n \cdot P} \frac{\bar n^{\mu}}{2} - \frac{1}{2} \Delta_{\perp}^{\mu} + \mathcal{O}\left(\frac{t}{Q},\frac{m^2}{Q}\right). \notag
\end{align}

\item[2.] The Kivel-Polyakov (KP) frame, where $P_{\perp} = q_{\perp} = 0$, adopted, for example, in \cite{Kivel:2000cn}. We have 
\begin{align}
\Delta_{\perp}^2 = 4m^2 \eta^2 + t(1- \eta^2)\,,
\end{align}
and
\begin{align} \notag
p^{\mu} &= (1+\eta) \bar n \cdot P \frac{n^{\mu}}{2} + (1-\eta) \frac{m^2 - t/4}{\bar n \cdot P} \frac{\bar n^{\mu}}{2} - \frac{1}{2} \Delta_{\perp}^{\mu},
\\ \notag
p'^{\mu} &= (1-\eta) \bar n \cdot P \frac{n^{\mu}}{2} + (1+\eta) \frac{m^2 - t/4}{\bar n \cdot P} \frac{\bar n^{\mu}}{2} + \frac{1}{2} \Delta_{\perp}^{\mu},
\\
q^{\mu} &= - (\rho + \eta) \bar n \cdot P \frac{n^{\mu}}{2} + \frac{Q^2}{\rho \bar n \cdot P} \frac{\bar n^{\mu}}{2} + \mathcal{O}\left(\frac{t}{Q},\frac{m^2}{Q}\right),
\\
q'^{\mu} &= - (\rho - \eta) \bar n \cdot P \frac{n^{\mu}}{2} + \frac{Q^2}{\rho \bar n \cdot P} \frac{\bar n^{\mu}}{2} - \Delta_{\perp}^{\mu} + \mathcal{O}\left(\frac{t}{Q},\frac{m^2}{Q}\right). \notag
\end{align}

\item[3.] The Braun-Manashov-Pirnay (BMP) frame \cite{Braun:2012hq}, where $\Delta_{\perp} = q_{\perp}' = 0$. Then
\begin{align}
P_{\perp}^2 = m^2 + \frac{t}{4} \frac{1-\eta^2}{\eta^2}\,,
\end{align}
and
\begin{align} \notag
p^{\mu} &= (1+\eta) \bar n \cdot P \frac{n^{\mu}}{2} + \frac{1-\eta}{4\eta^2} \frac{(-t)}{\bar n \cdot P} \frac{\bar n^{\mu}}{2} + P_{\perp}^{\mu},
\\ \notag
p'^{\mu} &= (1- \eta) \bar n \cdot P \frac{n^{\mu}}{2} + \frac{1+\eta}{4\eta^2} \frac{(-t)}{\bar n \cdot P} \frac{\bar n^{\mu}}{2} + P_{\perp}^{\mu},
\\
q^{\mu} &= - (\rho + \eta) \bar n \cdot P \frac{n^{\mu}}{2} + \frac{Q^2}{\rho \bar n \cdot P} \frac{\bar n^{\mu}}{2} + \mathcal{O}\left(\frac{t}{Q},\frac{m^2}{Q}\right),
\\ \notag
q'^{\mu} &= - (\rho - \eta) \bar n \cdot P \frac{n^{\mu}}{2} + \frac{Q^2}{\rho \bar n \cdot P} \frac{\bar n^{\mu}}{2}.
\end{align}
\end{itemize}
We will use almost exclusively the BMP frame, which is convenient since the photons have vanishing $\perp$ momentum at NLP accuracy. In the case of DVCS, $q'$ is exactly a light-like vector. 

Our derivation relies on the choice of the BMP frame, more precisely, on the condition $q_{\perp}' = q_{\perp} = 0$. Then,
given that the number and virtuality of momentum regions is a Lorentz invariant \cite{Pecjak:2005uh},
we conclude that if factorization holds in one frame, it also holds in another, at least as long as the scaling of the external momenta is left invariant. Furthermore, using the reparametrization invariance (RPI) \cite{Manohar:2002fd, Chay:2002vy} of the amplitude, which represents the residual Lorentz symmetry in SCET, we can convert a factorized expression from one frame to another. This is done in section \ref{sec: RPI}.


\section{Method of regions example}
\label{sec: example}

\begin{figure}
\centering
\includegraphics[scale=.5]{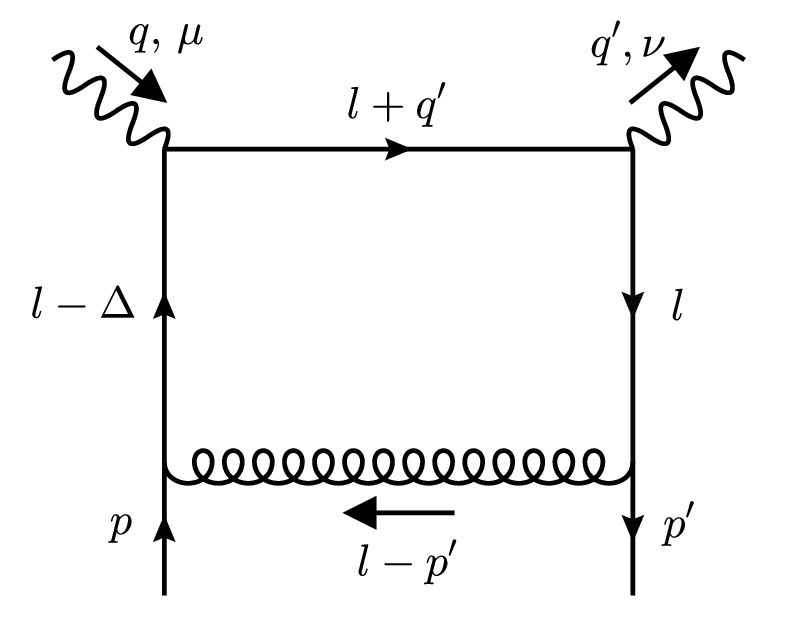}
\caption{Box diagram with DVCS kinematics.}
\label{fig: box dia}
\end{figure}

\begin{figure}
\centering
\includegraphics[scale=.19]{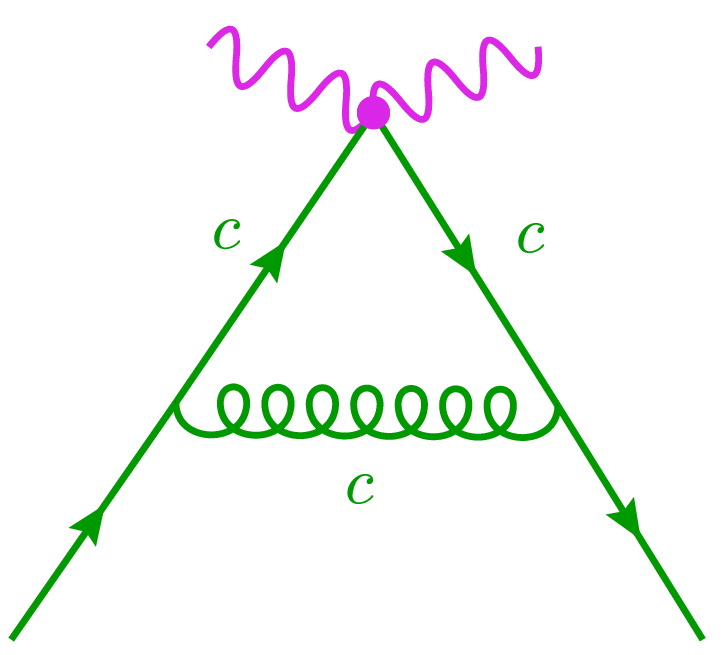} \hspace{.5cm}
\includegraphics[scale=.19]{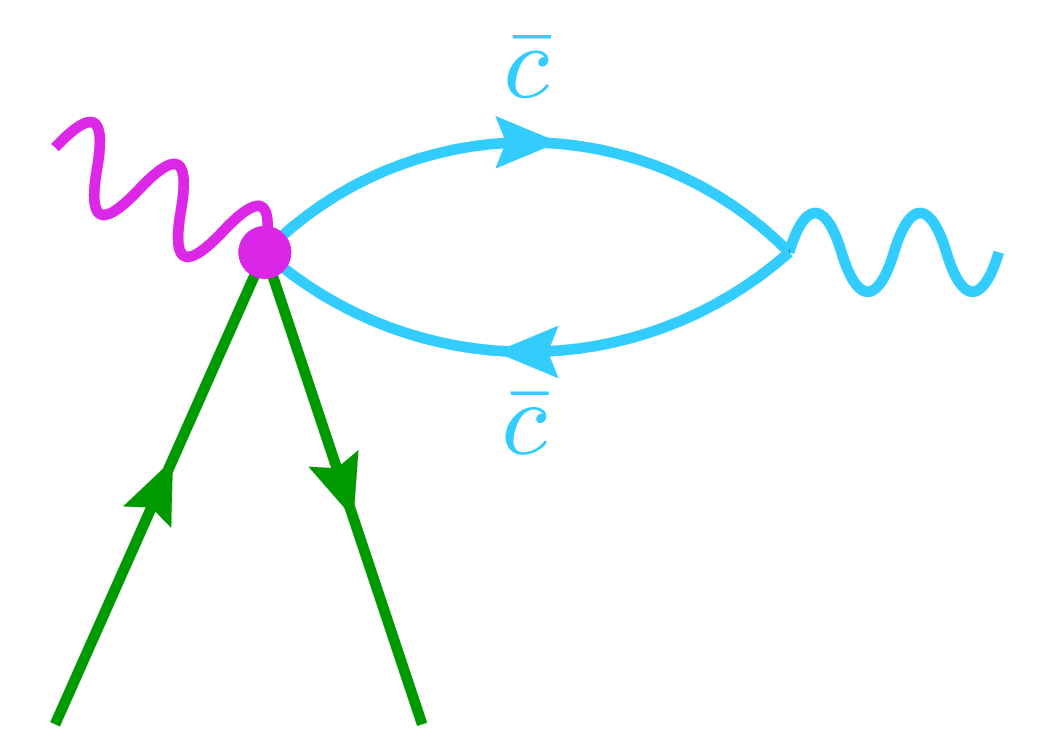}
\includegraphics[scale=.25]{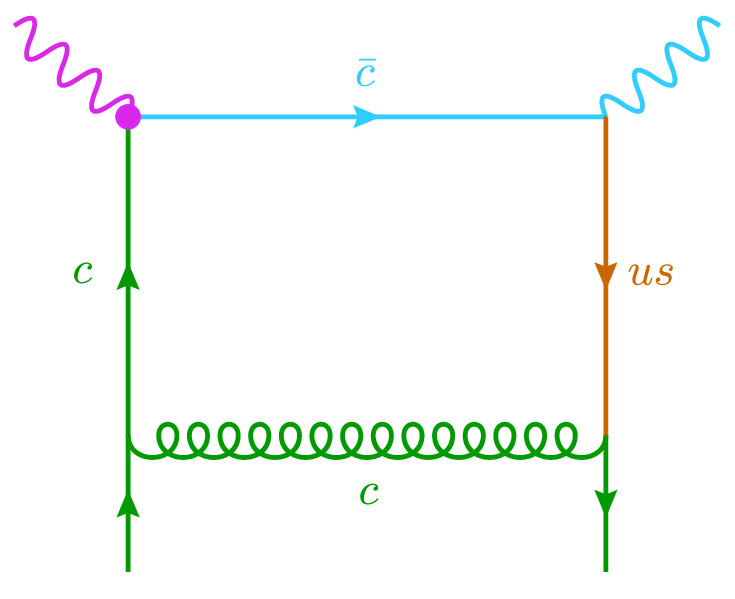}
\\
\hspace{-2cm} \quad (a) \hspace{3cm} \quad (b) \hspace{6cm} (c) \hspace{-1cm}
\caption{Reduced diagrams of figure \ref{fig: box dia} corresponding to the regions in eq. \eqref{eq: regions}.}
\label{fig: reduced}
\end{figure}

It is instructive to start with a ``simple'' example to substantiate the all-order arguments made later. We employ the method of regions \cite{Beneke:1997zp}, i.e.,  we assume various scalings\footnote{See \cite{Ma:2023hrt} for a recent discussion of the problem of identifying relevant regions.} of loop momentum with expansion parameter $\lambda$, perform expansion of the integrand in $\lambda$, integrate over the entire range. The sum of all the regions represents the expansion of the integral in $\lambda$. 

This section aims to elucidate what are the relevant IR regions.
In the case where both photons are far off-shell it is easy to see that the collinear region is the only relevant IR region at NLP.
Below, we consider the more interesting case of DVCS, i.e. $\rho = \eta$. We will find that for this one-loop example, it is indeed also true that only the collinear region contributes to NLP accuracy. 

We consider the box diagram in figure \ref{fig: box dia}, which reads (ignoring the trivial color algebra)
\begin{align}
I^{\mu \nu} &= \int d^dl \, \frac{\bar u(p') \gamma_{\alpha} \s l \gamma^{\nu} (\s l + \s q') \gamma^{\mu} (\s l - \s \Delta) \gamma^{\alpha} u(p)}{[(l+q')^2 + i0][ (l- \Delta)^2 + i0] [l^2 + i0] [(l-p')^2 + i0 ] }.
\end{align}
To get an integrand that is homogeneous in $\lambda$, we need to expand the spinor $u(p)$ in terms of the collinear spinor $u_c(p) = \frac{\s n \sbar n}{4} u(p)$. Using the Dirac equation $\s p u(p) = 0$, we get
\begin{align}
u(p) - u_c(p) = \frac{\sbar n \s n}{4} u(p) = -\frac{\sbar n \s p_{\perp}}{2 \bar n \cdot p} u_c(p),
\end{align}
so we have
\begin{align}
u(p) = \left ( 1 - \frac{\sbar n \s p_{\perp}}{2\bar n \cdot p}  \right ) u_c(p), \qquad
\bar u(p') = \bar u_c(p') \left ( 1 - \frac{\s p_{\perp}' \sbar n }{2 \bar n \cdot p'}  \right ).
\label{eq: u to un}
\end{align}
These relations are also critical in the matching procedure discussed in section \ref{sec: calc}.

First note that the hard region is trivially present and gives the bare contribution to the one-loop coefficient functions. Let us recall  that $\Delta, p,p' \sim p_c$ and $q' \sim p_{\bar c}$.
\begin{itemize}
\item Start with the collinear region: $l\sim (\lambda^2, 1 , \lambda)Q$. The reduced diagram is shown in figure~\ref{fig: reduced}(a):
\begin{align}
\text{collinear}: \qquad I_c \sim \underbrace{ \lambda^4 }_{\text{mom. space volume}} \, \underbrace{ \lambda^2 }_{\text{ numerator }} \, \underbrace{ \lambda^{-6}}_{\text{denominator}} \sim \lambda^0.
\end{align}
To see this, we note that each collinear denominator, except for the first one, gives a power of $\lambda^{-2}$. In the numerator, note that $\bar u(p') \gamma_{\perp \alpha} \s n,\; \s n \gamma_{\perp}^{\alpha} u(p) \sim \lambda$, so the large component of $\s l$ and $\s l - \s \Delta$ contributes only with suppressed spinor components. On the other hand, for $\s l + \s q'$, we can pick the $n \cdot q' \frac{\sbar n}{2}$ component, which scales as $\lambda^0$. Notice that in this region, we can replace, up to terms of $\f O(\lambda^2)$,\begin{align}
\frac{\s l + \s q'}{(l+q')^2 + i0} \longrightarrow \frac{1}{\bar n \cdot l + i0} \left ( \frac{\sbar n}{2} +  \frac{\s l_{\perp}}{n \cdot q'} \right ) \;.
\label{eq: hard propagator}
\end{align}
To get the factorization of the graph in the usual language, we would write $\bar n \cdot l = (x-\eta) \bar n \cdot P$. The first term in the brackets on the righthand side of eq. \eqref{eq: hard propagator} is then nothing but the contribution to the leading order coefficient function $\frac{1}{x - \eta + i0}$. The second term in the bracket contributes to the coefficient function of a twist-3 operator. 

\item Continue with the anti-collinear region: $l\sim ( 1, \lambda^2 , \lambda)Q$. Notice that in the strictly on-shell limit, this region leads to a scaleless integral because $q'^2 = 0$, i.e., there is no physical scale in the process corresponding to anti-collinear virtuality. Instead, let us consider the photon to be slightly off-shell $q'^2 \sim \lambda^2 Q$ by introducing a non-zero $\lambda^2 Q \sim \bar n \cdot q' \neq 0$. This off-shellness can be physically related to the breakdown of perturbative expansion below scale $\Lambda_{\rm QCD}$ for anti-collinear matrix elements and experimentally it can be thought of as a finite energy resolution of the photon detector. 
The reduced diagram is shown in figure \ref{fig: reduced}(b). Superficially:
\begin{align}
\text{anti-collinear}: \quad & I_{\bar c} \sim \underbrace{ \lambda^4 }_{\text{mom. space volume}} \, \underbrace{ \lambda^0 }_{\text{ numerator }} \, \underbrace{ \lambda^{-4}}_{\text{denominator}} \sim \lambda^0.
\end{align}
The power-counting for the numerator requires some explanation. In particular, it depends heavily on the polarization of photons. The largest term responsible for the leading power behaviour is obtained by taking the $\s n$ components of $\gamma^{\nu}$ and $\gamma^{\mu}$. Then, we can take the $\sbar n$ component of $\s l, \s l + \s q', \s l - \s \Delta$, giving the leading power. However, this contributes only to the unphysical longitudinal polarization of the outgoing photon. For physical photon polarization, we can effectively replace $\gamma^{\nu} \rightarrow \gamma_{\perp}^{\nu}$. Let us assume this from now on. 
In the $\bar n$-collinear region up to terms of $\f O(\lambda^2)$ the integral can be written as
\begin{align}
I_{\bar c} = \frac{\bar n^{\mu}}{\bar n \cdot \Delta \, \bar n \cdot p'} \int d^dl \, \frac{\bar u_c \gamma_{\perp\alpha} \Big [\s l_{\perp} \gamma_{\perp}^{\nu} n \cdot (l + q') n \cdot l \frac{\sbar n}{2} + n \cdot l \frac{\sbar n}{2} \gamma_{\perp}^{\nu} \s l_{\perp} \Big ] n \cdot l \gamma_{\perp}^{\alpha} u_c}{[(l+q')^2 + i0] (l^2 + i0) ( n \cdot l + i0) ( n\cdot l - i0 ) }.
\end{align}
Now, it is clear that $I_{\bar c}$ is proportional to $q_{\perp}'^{\nu} = 0$, so $I_{\bar c}$ vanishes. We conclude that the anti-collinear region is $\mathcal{O}(\lambda^2)$. We will find in section \ref{sec: nontarget regions} that this conclusion remains true to all orders.

\item Finally, consider the ultrasoft region $l \sim (\lambda^2, \lambda^2, \lambda^2) Q$. The reduced diagram is shown in figure \ref{fig: reduced}(c). Superficially:
\begin{align}
\text{ultrasoft}: \quad & I_{us} \sim \underbrace{ \lambda^8 }_{\text{mom. space volume}} \, \underbrace{ \lambda^3 }_{\text{ numerator }} \, \underbrace{ \lambda^{-10}}_{\text{denominator}} \sim \lambda^1.
\end{align}
The integral in the ultrasoft region can be written as
\begin{align}
I_{us} = \int d^dl\, \frac{\bar u_c \gamma_{\perp\alpha} \s l_{\perp} \gamma_{\perp}^{\nu} n \cdot q' \frac{\sbar n}{2} \gamma_{\perp}^{\mu} \bar n \cdot \Delta \frac{\s n}{2} \gamma_{\perp}^{\alpha} \frac{\sbar n \s P_{\perp}}{2\bar n \cdot p} u_c }{[n \cdot q' \, \bar n \cdot l + q'^2 + i0][ - \bar n \cdot \Delta \, n \cdot l + \Delta^2 + i0] [ - \bar n \cdot p' n \cdot l + p'^2 + i0] (l^2 + i0)}
\label{eq: Is mor}
\end{align}
Notice that we can drop $l_{\perp}$ in each denominator, except for $l^2 + i0$. Thus the $l_{\perp}$ integral is odd under $l_{\perp} \rightarrow -l_{\perp}$, so we get zero. Consequently, the ultrasoft region is $\mathcal{O}(\lambda^2)$. However this result should be interpreted with some care. Firstly such a kind of cancellation arises due to the chiral symmetry of massless quarks, which does not hold beyond perturbation theory.
Moreover, as we have mentioned before, the ultrasoft region is problematic when viewing $I^{\mu \nu}$ in the context of a physical process, where the collinear and ultrasoft lines are to be interpreted as corresponding to a universal non-perturbative function. This will be further discussed in section \ref{sec: nontarget regions}.
\end{itemize}
We remark that, strictly speaking, one also has to take into account the overlap contributions\footnote{The overlap contributions are also referred to as double-counting subtractions~\cite{Collins:2011zzd} or zero-bin subtractions~\cite{Manohar:2006nz} in various contexts. } as formulated in~\cite{Jantzen:2011nz} (see also \cite{Ferrera:2023vsw} for a recent discussion). However, it is easy to see that these give scaleless integrals here. 

This concludes the discussion of the regions that will be used later in the effective theory. Since we have found that $I_{us}, I_{\bar c} = \mathcal O(\lambda^2)$, we can expect (without any calculation) that the sum of the hard and the collinear regions reproduce the asymptotic expansion of $I^{\mu\nu}$ as $\lambda \rightarrow 0$ up to terms of $\mathcal O(\lambda^2)$. To definitely conclude this we need to argue that all other possible regions either give the same expansions as the hard or collinear region (in that case they can be considered as subsumed by the hard and collinear are automatically taken into account) or give scaleless or $\mathcal O(\lambda^2)$ contributions. We discuss the some interesting potential candidates in the following.
\begin{itemize}
    \item The soft region $l \sim (\lambda, \lambda, \lambda) Q$ gives a scaleless integral
\begin{align}
I_s = \int d^dl \, \frac{\bar u_c \gamma_{\perp\alpha} \s l_{\perp} \gamma_{\perp}^{\nu} n \cdot q' \frac{\sbar n}{2}\gamma_{\perp}^{\mu} \Big ( \s l_{\perp} \gamma_{\perp}^{\alpha} + \bar n \cdot \Delta \frac{\s n}{2} \gamma_{\perp}^{\alpha} \frac{\sbar n \s P_{\perp}}{2\bar n \cdot p}  \Big ) u_c}{[n \cdot q' \bar n \cdot l + i0] [- \bar n \cdot \Delta \, n \cdot l + i0 ] [- \bar n \cdot p' \, n \cdot l  + i0] (l^2 + i0)} = 0.
\label{eq: Is}
\end{align}
This is always true if there are only collinear and anti-collinear external particles, since $(p_s + p_c)^2 \sim \bar n \cdot p_c \, n \cdot p_s$. Hence, with respect to the asymptotic expansion of $I^{\mu \nu}$, the soft region is not needed. For future reference we also perform the power-counting
\begin{align}
\text{soft}: \quad & I_{s} \sim \underbrace{ \lambda^4 }_{\text{mom. space volume}} \, \underbrace{ \lambda^2 }_{\text{ numerator }} \, \underbrace{ \lambda^{-5}}_{\text{denominator}} \sim \lambda^1.
\label{eq: soft pc}
\end{align}
It is worth noting that a soft region can emerge as a secondary effect of rapidity regularization. In some exclusive processes, collinear and anti-collinear regions are ill-defined under pure dimensional regularization, necessitating the use of additional rapidity-type regulators. These regulators can, in turn, introduce non-vanishing soft contributions. While rapidity divergences are absent in our current scenario, their presence in more complex exclusive processes cannot be ruled out. 
\end{itemize}
There exist arguments \cite{Ma:2023hrt} stating that the regions that we have discussed are indeed sufficient and there are no ``exotic'' regions that need to be considered in some generality. However, the arguments in \cite{Ma:2023hrt} apply only to wide-angle scattering, where there are no small scalar products between two collinear external momenta such as $\Delta \cdot p' \sim \lambda^2$ and $\text{sgn}(\bar n \cdot \Delta) \neq \text{sgn}(\bar n \cdot p')$. This pinches the $n \cdot l$ component to be $\f O(\lambda^2)$ while the $l^2$ denominator being far off-shell. 
The only special case that can appear due to this subtlety is the Glauber region, which we will discuss just now.
\begin{itemize}
\item The Glauber region $l \sim (\lambda^2, \lambda^2, \lambda )Q$ gives
\begin{align}\notag
I_G &= \int d^dl \, \frac{(...)}{[l_{\perp}^2 + \bar n \cdot l \, n \cdot q' + 2l_{\perp} \cdot q_{\perp}' + q'^2 + i0][l_{\perp}^2 + i0] }
\\
&\quad \times \frac{1}{[l_{\perp}^2 - n \cdot l \, \bar n \cdot \Delta - 2l_{\perp} \cdot \Delta_{\perp} + \Delta^2 + i0] [l_{\perp}^2 - n \cdot l \, \bar n \cdot p' - 2l_{\perp} \cdot p_{\perp}' + p'^2 + i0]}.
\end{align}
It superficially gives a leading power contribution
\begin{align}
\text{Glauber}: \quad & I_G \sim \underbrace{ \lambda^6 }_{\text{mom. space volume}} \, \underbrace{ \lambda^2 }_{\text{ numerator }} \, \underbrace{ \lambda^{-8}}_{\text{denominator}} \sim \lambda^0.
\label{eq: Glauber pc}
\end{align}
Note that the $\bar n \cdot l$ integral actually diverges at infinity. This is typical for Glauber regions. Strictly speaking this divergence is cancelled by correctly taking into account the overlap subtractions. More conveniently one can introduce a rapidity regulator $(l+q')^2 \rightarrow [(l+q')^2]^{1+\alpha}$ with $\alpha > 0$, which allows from a deformation of the $\bar n \cdot l$ contour to infinity making the integral vanish. In the language of \cite{Collins:2011zzd} one can argue that $\bar n \cdot l$ is not pinched to be of $\mathcal O(\lambda^2)$. This is obvious, since there is only a single pole (corresponding to the denominator $(l+q')^2 + i0$) that is close to the origin $\bar n \cdot l = 0$. The $\bar n \cdot l$ contour can therefore be deformed to such that $|\bar n \cdot l| \sim Q$, so that $l \sim p_c$.
\end{itemize}


We conclude that as long as the real photon is transversely polarized and $q_{\perp}' = 0$, the only relevant non-hard contribution to $I$ at NLP is the collinear contribution. We will find that the operatorial expansion in the SCET framework readily generalizes these arguments to all graphs and all orders. 
However, the conclusions for the soft and ultrasoft regions for the simple one-loop graph $I$ should not be extended to the non-perturbative domain. We will see that the argument used to conclude that the ultrasoft region vanishes relies on chiral symmetry, which is spontaneously broken by non-perturbative effects.

\section{All-order factorization}
\label{sec: factorization}

\subsection{Operator building blocks}
\label{sec: operator building blocks}
\begin{figure}
\centering
\includegraphics[scale=.3]{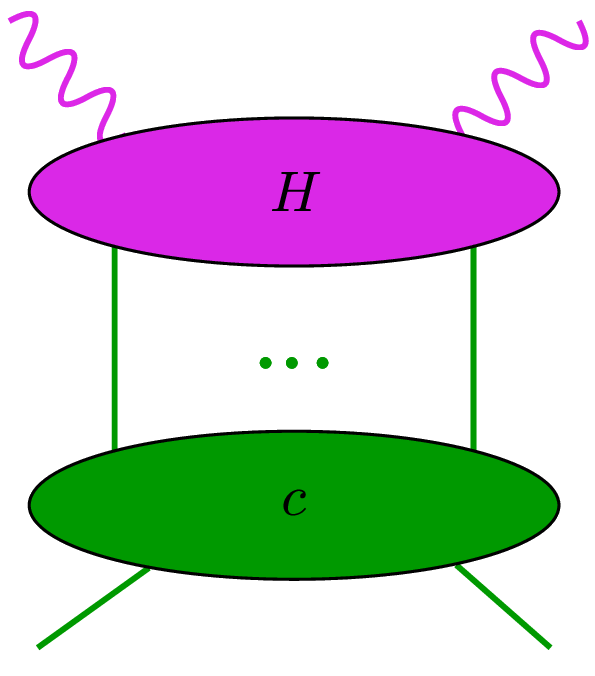} \hspace{.5cm}
\includegraphics[scale=.3]{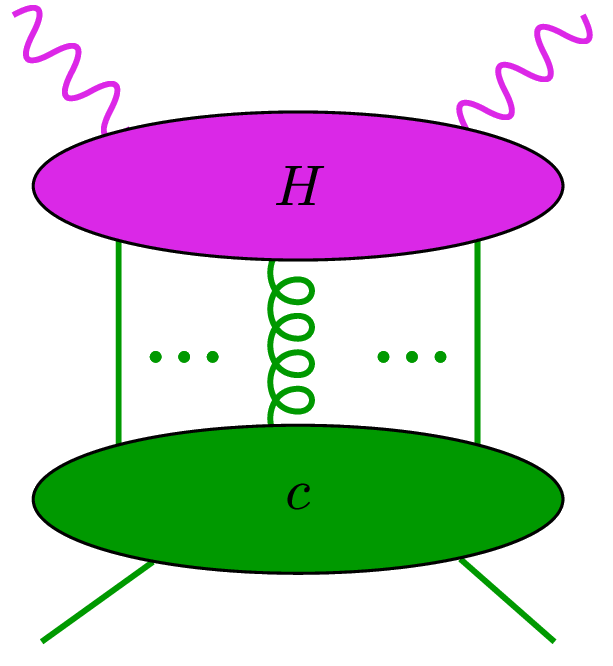}
\\
\hspace{-.5cm} \quad (a) \hspace{4.5cm} \quad (b)
\caption{All-order reduced diagrams involving only collinear regions. The dots denote an arbitrary number of scalar-polarized gluons. The quark lines connecting the hard and collinear subgraph can also be transversely polarized gluon lines. The right-hand diagram (b) contains an additional transversely polarized gluon connecting the hard and collinear subgraph, corresponding to an NLP correction. }
\label{fig: allorderreduced 1 and 4}
\end{figure}
This section relies on the definitions introduced in Appendix \ref{sec: SCETI L}.
To derive factorization using the EFT formalism, we first have to identify all gauge-invariant operators, and we can use reparametrization invariance to constrain the operator basis.

Collinear gauge invariance is automatically satisfied by the introduction of collinear gauge invariant collinear building blocks:
\begin{equation}
 \begin{array}{ll}
\displaystyle 
\chi_c(x)=W_c^{\dagger}(x)\xi_c(x) \sim \lambda^1  & {\quad{\text{collinear quark,}}} \\[0.3cm]
\displaystyle 
\mathcal{A}^{\mu}_{c\perp}(x)= 
W_c^{\dagger}(x)\big[iD^{\mu}_{c\perp}W_c(x)\big] \sim \lambda^1
 & {\quad{\text{collinear gluon,}}}
\end{array}
\end{equation}
which are composed of the $n$-collinear fields \textit{before} the ultrasoft decoupling transformation
\begin{align}
\xi_c \rightarrow Y_n^{[n]} \xi_c, \qquad A_c \rightarrow Y_n^{[n]} A_c Y_n^{[n] \dagger}.
\label{eq: decoupling trafo}
\end{align}
Further, the subleading gluon building block $n \cdot \mathcal{A}_{{n}}(x)= 
W_c^{\dagger}(x)\big[i n \cdot D_{c}W_c(x)\big]$, can be eliminated using equation of motion \cite{Beneke:2017ztn}. 
The collinear building blocks can be acted upon with derivative operators 
\begin{align}
   \partial_\perp \sim \lambda^1, \qquad n \cdot D_{us} \sim \lambda^2,
\end{align}
where, due to the multipole expansion of the ultrasoft fields, only the $n \cdot D_{us}$ has to be a ultrasoft gauge covariant derivative rather than an ordinary derivative. This derivative operator can also be eliminated from operator basis \cite{Beneke:2017ztn,Beneke:2019kgv}. 
 
In addition, the operators may contain ultrasoft building blocks
\begin{equation}
 \begin{array}{ll}
\displaystyle 
q_{us}(0) \sim \lambda^3 & {\quad{\text{ultrasoft quark,}}} \\[0.3cm]
\displaystyle 
F_{us}^{\mu\nu}(0) \sim \lambda^4
 & {\quad{\text{ultrasoft gluon,}}}
\end{array}
\end{equation}
which only start contributing from relative $\mathcal{O}(\lambda^3)$ order \cite{Beneke:2017ztn}. 
The ultrasoft gauge invariance of a generic operator is equivalent to the color neutrality of the entire operator.

After the ultrasoft decoupling transformation, the operators may contain the ultrasoft Wilson lines that sum up the unsuppressed interactions of scalar-polarized ultrasoft gluons with the collinear fields.  If we consider an operator containing only collinear building blocks (understood to be evaluated on the $\bar n$-light-cone) sourced by a color singlet, the ultrasoft Wilson lines cancel.  

\subsection{Only collinear modes}
\label{sec: only ncoll}

Let us focus only on operators constructed from $n$-collinear building blocks. 
The leading power operators that have non-zero overlap with the off-forward hadron matrix element $\bra {p'} ... \ket p$ contain two building blocks:
\begin{align}
\f O_{\bar \chi \chi}^{[\Gamma]}(s_1,s_2) = 
\bar \chi_c(s_1 \bar{n}) \frac{\Gamma}{2} 
\chi_c(s_2 \bar{n}), 
\qquad \f O_{\f A \f A}^{\mu_1 \mu_2}(s_1,s_2) = \text{tr}( \f A_{c\perp}^{\mu_1}(s_1 \bar{n}) \f A_{c\perp}^{\mu_2}(s_2\bar{n})).
\label{eq: LP operators}
\end{align}
As usual, the SCET operators are non-local along the lightcone. 
The hadronic matrix element of these operators can be identified with the lower $n$-collinear subgraph of the reduced graph figure~\ref{fig: allorderreduced 1 and 4}(a).
As for the basis of Dirac matrices one conventionally chooses $\Gamma \in \{ \sbar n, \sbar n\gamma_5, \sigma^{\mu\nu}_{~\, \perp} \bar n_{\mu} \}$. We do not decompose the Lorentz indices of the operators $\f O_{\f A \f A}^{\mu_1 \mu_2}$ since gluons do not appear at leading order in $\alpha_s$ and we will only consider $\alpha_s^0$ matching. We only note that there exist three possible tensor structures; see~\cite{Ji:2023xzk} for the general decomposition in $d$ dimensions.

The next-to-leading power color-singlet operators are
\begin{align} \notag
\f O_{\bar \chi \p \chi}^{[\Gamma] \mu}(s_1,s_2) &= \bar \chi_c(s_1 \bar n) \frac{\Gamma}{2} i\overleftrightarrow{\p}_{\perp}^{\mu} \chi_c(s_2 \bar n), 
\\ \notag
\f O_{\p (\bar \chi \chi)}^{[\Gamma] \mu}(s_1,s_2) &= i\p_{\perp}^{\mu} (\bar \chi_c(s_1 \bar n) \frac{\Gamma}{2} \chi_c(s_2 \bar n) ), 
\\ \notag
\f O_{\bar \chi \f A \chi}^{[\Gamma] \mu}(s_1,s_2,s_3) &= \bar \chi_c(s_1 \bar n) \frac{\Gamma}{2} \f A_{c\perp}^{\mu}(s_2 \bar n) \chi_c(s_3 \bar n),
\\ 
\f O_{\f A \p \f A}^{\mu_1 \mu_2 \mu_3}(s_1,s_2) &= \text{tr}( \f A_{c\perp}^{\mu_1}(s_1 \bar n) i\overleftrightarrow{\p}_{\perp}^{\mu_2} \f A_{c\perp}^{\mu_3}(s_2 \bar n) ), \label{eq: NLP operators}
\\ \notag
\f O_{\p(\f A \f A)}^{\mu_1 \mu_2 \mu_3}(s_1,s_2) &= \p_{\perp}^{\mu_1} \text{tr}( \f A_{c\perp}^{\mu_2}(s_1 \bar n) \f A_{c\perp}^{\mu_3}(s_2 \bar n) ),
\\ \notag
\f O_{\f A \f A \f A}^{\mu_1 \mu_2 \mu_3}(s_1,s_2,s_3) &= \text{tr}( \f A_{c\perp}^{\mu_1}(s_1 \bar n) \f A_{c\perp}^{\mu_2}(s_2 \bar n) \f A_{c\perp}^{\mu_3}(s_3 \bar n) ).
\end{align}
Note that $\f O_{\bar \chi \p \chi}$ and $\f O_{\p (\bar \chi \chi)}$ correspond to the reduced graph in figure~\ref{fig: allorderreduced 1 and 4}(a), where we keep the $\mathcal{O}(\lambda^1)$ term of the hard subgraph. On the other hand, the tri-local operators correspond to the reduced graph in figure~\ref{fig: allorderreduced 1 and 4}(b). It can be readily checked that the SCET power-counting agrees with the graphical counting rules~\cite{Collins:2011zzd}, upon noting that the external hadron collinear states count for $\lambda^{-1}$ each. 

Note that $\f O_{\p(\bar \chi \chi)}^{[\Gamma]\mu}$ and $\f O_{\p(\f A \f A)}^{[\Gamma]\mu}$ are trivially related to leading power operators, e.g.
\begin{align}
\bra{p'} \f O_{\p(\bar \chi \chi)}^{[\Gamma]\mu} \ket p = \bra{p'}  \left [-i{\rm P}_{\perp}^{\mu} , \f O_{\bar \chi \chi}^{[\Gamma]} \right ]  \ket p = -i \Delta_{\perp}^{\mu} \bra{p'} \f O_{\bar \chi \chi}^{[\Gamma]\mu} \ket p,
\end{align}
where ${\rm P}_{\perp}^{\mu}$ is the momentum operator. In the BMP frame, these operators do not contribute. 

To define parton density functions, it is first convenient to eliminate one of the positions by a translational invariance in the collinear sector. The (SCET) parton densities are then defined as Fourier-transformed matrix elements with respect to the relative light-cone positions $s_i$.  We use the following  convention for bi-local operators:
\begin{align}
\bra{p'} \f O_{\bar \chi \chi}^{[\Gamma]}(-s/2, s/2) \ket p &= \bar n \cdot P \intR dx\, e^{-ixs \bar n \cdot P} F_{\bar \chi \chi}^{[\Gamma]}(x),
\\
\bra{p'} \f O_{\bar \chi \p \chi}^{[\Gamma]\mu}(-s/2,s/2 )\ket{p} &= (\bar n \cdot P)^2 \intR dx \, e^{-ixs\bar n \cdot P} F_{\bar \chi \p \chi}^{[\Gamma]\mu}(x).
\end{align}
In addition to momentum fraction $x$, the GPDs depend implicitly on $\eta, t$ and the factorization scale $\mu_F$. This dependence is tacitly implied and will be omitted from notation throughout.

For the quark-gluon-quark operator, we define
\begin{align} \notag
&\bra{p'} \f O_{\bar \chi \f A \chi}^{[\Gamma]\mu}(-s_1/2,s_2/2, s_1/2) \ket p 
\\
&=  (\bar n \cdot P)^2 \intR dx_1 dx_2 \, e^{-ix_1 s_1 \bar n \cdot P -i x_2 s_2 \bar n \cdot P} F_{\bar \chi \f A \chi}^{[\Gamma]\mu}(x_1,x_2)
\end{align}
and similarly for the other tri-local operators.

If the operators in eqs. \eqref{eq: LP operators} and \eqref{eq: NLP operators} are indeed sufficient at NLP, the factorization follows immediately
\begin{align} \notag
T^{\mu \nu} &= \sum_{\ell} \intR \frac{ds}{2\pi} \, \hat C_{\ell}^{\mu \nu}(s) \bra{p'} \f O_{\ell}(-s/2, s/2) \ket p 
\\
&\quad + \sum_{\ell} \intR \frac{ds_1 ds_2}{(2\pi)^2} \, \hat C_{\ell}^{\mu \nu}(s_1,s_2) \bra {p'} \f O_{\ell}(-s_1/2, s_2/2, s_1 /2) \ket p + \f O(\lambda^2),
\end{align}
where $\ell$ is a multiindex that runs over all operators eqs. \eqref{eq: LP operators} and \eqref{eq: NLP operators}. In terms of the Fourier transformed coefficient functions
\begin{align} \notag
C_{\ell}^{\mu \nu}(x) &= (\bar n \cdot P)^{\text{dim}\,\f O_{\ell} - 2}\intR \frac{ds}{2\pi} \, \hat C_{\ell}^{\mu \nu}(s) e^{-ixs\bar n \cdot P},
\\
C_{\ell}^{\mu \nu}(x_1,x_2) &= (\bar n \cdot P)^{\text{dim}\,\f O_{\ell} - 2} \intR \frac{ds_1 ds_2}{(2\pi)^2} \, \hat C_{\ell}^{\mu \nu}(s_1,s_2) e^{-i x_1 s_1 \bar n \cdot P - i x_2 s_2 \bar n \cdot P},
\end{align}
the factorization theorem reads
\begin{align}
T^{\mu \nu} = \sum_{\ell} \intR dx \, C_{\ell}(x) F_{\ell}(x) + \sum_{\ell} \intR dx_1 dx_2 \, C_{\ell}(x_1,x_2) F_{\ell}(x_1,x_2) + \f O(\lambda^2).
\label{eq: factorization theorem}
\end{align}
This factorization applies without caveats if both photons have hard virtualities. In that case, no anti-collinear modes contribute, and any ultrasoft interactions are absent. 

We remark that for light-cone distributions like the $F_{\ell}$'s one can \cite{Jaffe:1983hp,Diehl:1998sm} typically drop the time-ordering prescription in the matrix elements and assume compact support, supposedly contained in the region $-1 \leq x_j \leq 1$. Assuming this remains true in SCET, one can limit the range of integration in eq. \eqref{eq: factorization theorem} to $-1 \leq x_j \leq 1$, but we will never use this property.


\subsection{Anti-collinear and (ultra)soft contributions}
\label{sec: nontarget regions}
\begin{figure}
\centering
\includegraphics[scale=.33]{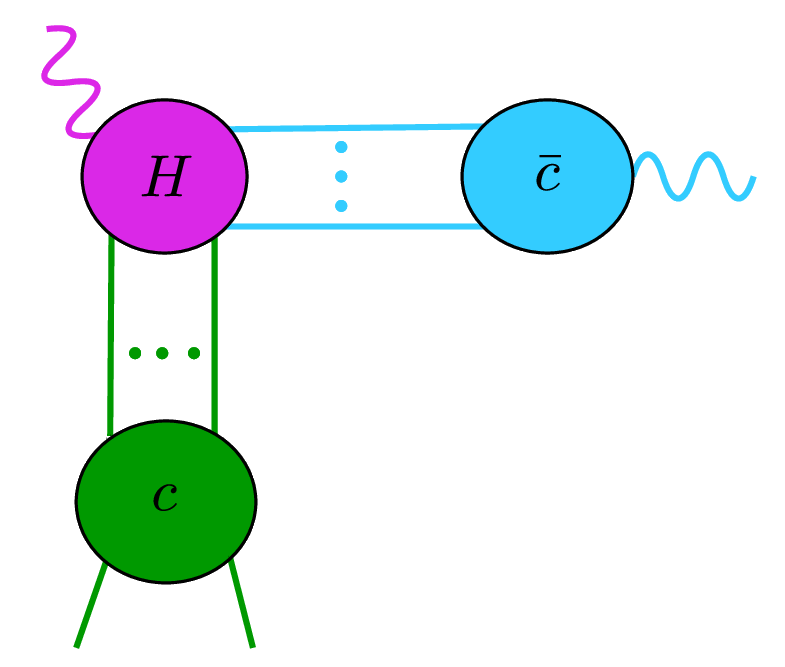} \hspace{.5cm}
\includegraphics[scale=.3]{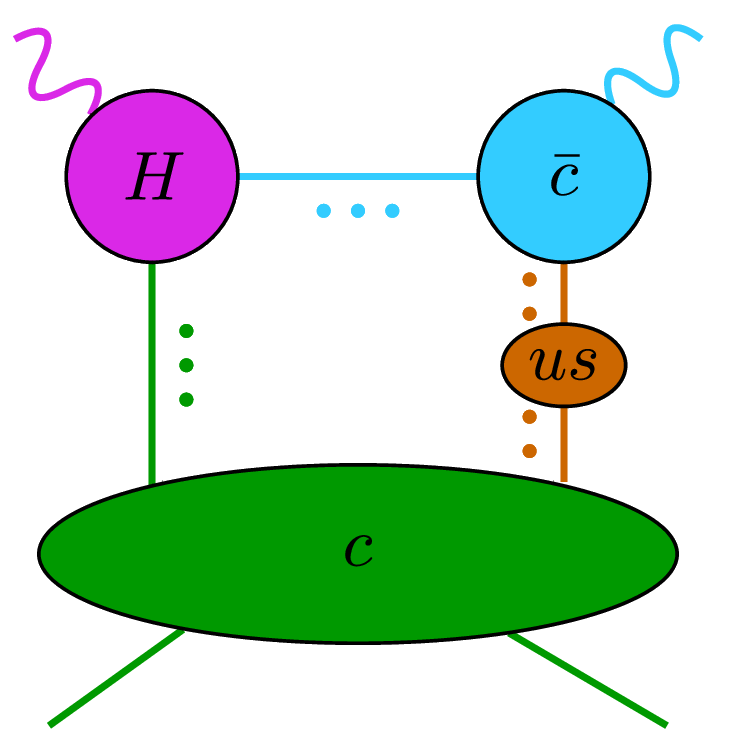}
\\
\hspace{-2cm} \quad (a) \hspace{8cm} \quad (b)
\caption{Reduced diagrams that involve anti-collinear and ultrasoft regions. The dots denote an arbitrary number of scalar-polarized gluons.}
\label{fig: allorderreduced 2 and 3}
\end{figure}

When one photon is on-shell, and therefore light-like, say in the anti-collinear direction, we have to take into account $\f L_{\bar c}^{(0)}$ and ultrasoft interactions, which are due to the subleading SCET Lagrangians $\f L_{c us}^{(1)}$ and $\f L_{\bar c us}^{(1)}$. For definiteness, we consider the case where the outgoing photon is on-shell $\rho = \eta$, but the arguments apply similarly to the case where the incoming photon is on-shell $\rho = - \eta$.

We first need to identify the QED interaction terms (in terms of SCET fields) that can mediate the emission of the real photon. We strictly work to leading order in the electromagnetic coupling, so we always consider the external photon amputated and the polarization vector implicitly contracted with the index $\nu$. After the ultrasoft decoupling transformation eq. \eqref{eq: decoupling trafo}, the corresponding interaction vertices in SCET are given by
\begin{align} \notag
\f T_{\bar c \bar c}^{\nu} &= i \int d^4z\, \bar \xi_{\bar c} \frac{\s n}{2} \Big [  \bar n^{\nu} e^{i q' \cdot z} + i \s D_{\bar c\perp} \frac{1}{in \cdot D_{\bar c}} \gamma_{\perp}^{\nu} e^{i q' \cdot z} + \gamma_{\perp}^{\nu} e^{i q' \cdot z} \frac{1}{i n\cdot D_{\bar c}} i\s D_{\bar c\perp}  
\\
&\hspace{2.3cm} - i\s D_{\bar c \perp}  \frac{1}{i n \cdot D_{\bar c}} n^{\nu} e^{iq' \cdot z} \frac{1}{i n \cdot D_{\bar c}} i \s D_{\bar c\perp} \Big ] \xi_{\bar c},
\label{eq: Tbarcbarc}
\\
\f T_{\bar c us}^{\nu} &= i \int d^4z\, \left [ \bar q_{us}^{[\bar n]} Y_{\bar n}^{[\bar n]}\gamma^{\nu} e^{iq' \cdot z} \chi_{\bar c} + \bar \chi_{\bar c} \gamma^{\nu} e^{iq' \cdot z} Y_{\bar n}^{[\bar n]} q_{us}^{[\bar n]} \right ].
\label{eq: Tbarcus}
\end{align}
Note that the components of $\f T_{\bar c \bar c}^{\nu}$ are not all of the same size because we amputated the photon field. In fact
\begin{align} \notag
n \cdot \f T_{\bar c \bar c} &= i \int d^4z \, \bar \xi_{\bar c} \s n e^{iq' \cdot z} \xi_{\bar c} \sim \lambda^{-2}, 
\\
\f T_{\bar c \bar c \perp}^{\nu} &= i \int d^4z\, \bar \xi_{\bar c} \frac{\s n}{2} \left [ i \s D_{\bar c\perp} \frac{1}{in \cdot D_{\bar c}} \gamma_{\perp}^{\nu} e^{i q' \cdot z} + \gamma_{\perp}^{\nu} e^{i q' \cdot z} \frac{1}{i n\cdot D_{\bar c}} i\s D_{\bar c\perp} \right ] \xi_{\bar c} \sim \lambda^{-1},
\\ \notag
\bar n \cdot \f T_{\bar c \bar c} &= -i \int d^4z\, \bar \xi_{\bar c} \s D_{\bar c\perp}  \frac{1}{i n \cdot D_{\bar c}} n^{\nu} e^{iq' \cdot z} \frac{1}{i n \cdot D_{\bar c}} i \s D_{\bar c\perp} \xi_{\bar c} \sim \lambda^0.
\end{align}
Compactly written, we have
\begin{align}
\f T_{\bar c \bar c}^{\nu} \sim p_{\bar c}^{\nu} \times \lambda^{-2}.
\end{align}
Moreover, $\bar n \cdot \f T_{\bar c us} = 0$ and $n \cdot \f T_{\bar c us} \sim \f T_{\bar c us  \perp}^{\nu} \sim \lambda^0$. 

First, we argue that power-suppressed currents in the anti-collinear directions do not contribute. Let us consider the hard current
\begin{align}
\f O_{\bar c \bar c c c }^{a_1 a_2 a_3 a_4} = \bar \chi_{\bar c}^{a_1} \chi_{\bar c}^{a_2} \bar \chi_c^{a_3} \chi_c^{a_4},
\end{align}
where $a_j$ are multiindices of spinor and color components. We have omitted ultrasoft Wilson lines from the decoupling transformation, which will drop out when taking matrix elements and using the color singlet properties of the collinear and anti-collinear matrix elements. Perform the power counting
\begin{align}
\underbrace{\bra {p'}}_{\sim \lambda^{-1}} \underbrace{ \f T_{\bar c \bar c}^{\nu} }_{\sim p_{\bar c}^{\nu} \times \lambda^{-2}} \underbrace{ \f O_{\bar c \bar c c c} }_{\sim \lambda^4} \underbrace{ \ket p }_{\lambda^{-1}} \sim p_{\bar c}^{\nu} \times \lambda^0.
\label{eq: T bar n n O}
\end{align}
This matrix element corresponds to the reduced graph in figure \ref{fig: allorderreduced 2 and 3}(a). 

The leading contribution, which is proportional to $\bar n^{\nu}$, can be readily identified with the method of regions treatment of the box diagram. Since we restrict ourselves to physical polarizations $\epsilon(q')^{\nu} = \epsilon(q')_{\perp}^{\nu}$ of the outgoing photons this contribution vanishes. 
There remains, however, an NLP contribution, which is proportional to the matrix element
\begin{align}
\bra{0} \f T_{\bar c \bar c \perp}^{\nu} \bar \chi_{\bar c }^{a_1} \chi_{\bar c}^{a_2} \ket 0 \sim \lambda^1.
\label{eq: barc ME}
\end{align}
To investigate this contribution, consider the possible twist-2 projections of the Dirac indices of the two quark fields. This gives three possibilities for $\Gamma$ in $\bra{0} \f T_{\bar c \bar c \perp}^{\nu} \bar \chi_{\bar c } \Gamma \chi_{\bar c} \ket 0$,
which are $\Gamma \in \{ \s n, \s n \gamma_5, n_{\mu} \sigma^{\mu j} \}$ ($i,j \in \{ 1,2 \}$ denote transverse indices). For the first two, i.e. $\{\s n, \s n \gamma_5\}$, the matrix element will be proportional to $q_{\perp}'^{\nu}$, which is zero in the BMP frame. Indeed, the only non-zero possibility is $\Gamma = n_{\mu} \sigma^{\mu j}$ and the anti-collinear matrix element in eq. \eqref{eq: barc ME} is then proportional to the photon distribution amplitude $\phi_{\gamma}$ \cite{Ioffe:1983ju, Balitsky:1989ry, Ball:2002ps}. By helicity conservation in the hard scattering the collinear matrix element is proportional to the transversity GPD $F_{\bar \chi \chi}^{[\bar n_{\mu} \sigma^{\mu j}]}$.
Thus, one might conclude that there is a corresponding non-zero contribution to the DVCS amplitude at NLP, which is proportional to
\begin{align}
\int du dx\, C(u,x) \, F_{\bar \chi \chi}^{[\bar n_{\mu} \sigma^{\mu j}]}(x) \phi_{\gamma}(u).
\end{align}
However, it can be shown that the corresponding hard coefficient function $C(u,x)$ vanishes in $d = 4$ to all orders in perturbation theory \cite{Diehl:1998pd, Collins:1999un}. We repeat this short argument in the following. First, note that the hard subamplitude, which is given by the blob $H$ in figure \ref{fig: allorderreduced 2 and 3}(a), can be written as
\begin{align}
C(u,x) = \omega_{ab} \text{tr} \{ \Gamma^a n_{\mu} \sigma^{\mu i} \Gamma^b \bar n_{\nu} \sigma^{\nu j} \},
\end{align}
where $\Gamma^a, \Gamma^b \in \{ 1, \gamma_5, \gamma^{\mu}, \gamma^{\mu} \gamma_5, \sigma^{\mu \nu} \} $ and $\omega_{ab}$ are the corresponding coefficients. $\Gamma^a$ and $\Gamma^b$ describe the two fermion lines that connect the hadron to the photon respectively. Since the chirality of the quarks is conserved during the hard scattering process they can each be written as a product of an odd number of $\gamma$ matrices. Hence, we have in fact only $\Gamma^a, \Gamma^b \in \{ \gamma^{\mu}, \gamma^{\mu} \gamma_5\} $ to consider. 
There are therefore only two cases to check, which are
\begin{align}
&\omega_{\alpha \beta} \text{tr}(\gamma^{\alpha} n_{\mu} \sigma^{\mu i} \gamma^{\beta} \bar n_{\nu} \sigma^{\nu j}),
\label{eq: tr1}
\\
&\omega_{\alpha \beta} \text{tr}(\gamma^{\alpha} n_{\mu} \sigma^{\mu i}\gamma^{\beta} \bar n_{\nu} \sigma^{\nu j} \gamma_5).
\label{eq: tr2}
\end{align}
The tensor $\omega^{\alpha \beta}$ is built out of $n^{\alpha (\beta)}, \bar n^{\alpha(\beta)}, \varepsilon_{\perp}^{\alpha \beta}$ and $g_{\perp}^{\alpha \beta}$ with Lorentz invariant coefficients. Any component proportional to $n^{\alpha(\beta)}$, $\bar n^{\alpha(\beta)}$ vanishes obviously in eqs. \eqref{eq: tr1} and \eqref{eq: tr2}, so the only remaining possiblities are $\omega_{\alpha \beta} \propto g_{\perp \alpha \beta}, \varepsilon_{\perp \alpha \beta}$. Now, note that 
\begin{align}
\gamma_{\perp}^{\alpha} n_{\mu} \sigma^{\mu i} \gamma_{\perp \alpha} = \varepsilon_{\perp \alpha \beta}\gamma_{\perp}^{\alpha} n_{\mu} \sigma^{\mu i} \gamma_{\perp}^{\beta} = 0, \qquad (d=4).
\end{align}
We therefore conclude that the contribution from $\f O_{\bar c \bar c c c}$ is $\f O(\lambda^2)$.
This is consistent with general observations about the absence of mixing of bi-local operators into local operators for massless theory \cite{Beneke:2017ztn,Beneke:2018rbh,Beneke:2019slt}. 

Next, consider the hard current
\begin{align}
\f O_{\bar c c}^{\mu}(s_1,s_2) = \bar \chi_{\bar c}(s_1 n) \gamma_{\perp}^{\mu} Y_{\bar n}^{\dagger}(0) Y_n(0) \chi_c(s_2 \bar n),
\label{eq: Obarcc}
\end{align}
where the product of the ultrasoft Wilson lines $Y_{\bar n}^{\dagger}(0) Y_n(0)$ originated from the decoupling transformation eq.~\eqref{eq: decoupling trafo}. Note that separate translation invariance in collinear and anti-collinear sectors allows us to set $s_1 = s_2 = 0$, i.e., $\f O_{\bar c c}$ becomes local at the position of the hard interaction vertex.
There are two superficially NLP contributions
\begin{align}
&\underbrace{ \bra{p'} }_{\sim \lambda^{-1}} \underbrace{ \f O_{\bar c c}^{\mu} }_{\sim \lambda^2} \underbrace{ \f T_{\bar c \bar c}^{\nu} }_{\sim p_{\bar c}^{\nu} \lambda^{-2}} \underbrace{ \f L^{(1)}_{\bar q_{us} A_{\bar c} \xi_{\bar c}}}_{\sim \lambda^1} \underbrace{ \f L^{(1)}_{\bar \xi_c A_c q_{us}} }_{\sim \lambda^{1}}  \underbrace{ \ket p }_{\sim \lambda^{-1}}  \sim p_{\bar c}^{\nu} \times \lambda^0, \notag
\\
&\underbrace{ \bra{p'} }_{\sim \lambda^{-1}} \underbrace{ \f O_{\bar c c}^{\mu} }_{\sim \lambda^2} \underbrace{ \f T_{\bar c us}^{\nu}}_{\sim \lambda^0} \underbrace{ \f L^{(1)}_{\bar \xi_c A_c q_{us}} }_{\sim \lambda^1} \underbrace{\ket p }_{\sim \lambda^{-1}} \sim \lambda^1,
\label{eq: usoft melts}
\end{align}
where
\begin{align}
\f L_{\bar q_{us} A_{\bar c} \xi_{\bar c} }^{(1)} &= i \int d^4z\, \bar q_{us}^{[\bar n]} Y_{\bar n}^{[\bar n]} W_{\bar c}^{\dagger} i \s D_{\bar c\perp} \xi_{\bar c}, \qquad \f L_{\bar \xi_c A_c q_{us}}^{(1)} = -i \int d^4z\, \bar \xi_c i \overleftarrow{\s D}_{c\perp} W_c Y_n^{[n] \dagger} q_{us}^{[n]}
\end{align}
are part of the subleading Lagrangians $\f L_{c us}^{(1)}$ and $\f L_{\bar c us}^{(1)}$, see Appendix \ref{sec: SCETI L}. The contributions from the matrix elements in eq. \eqref{eq: usoft melts} can be identified with the reduced graph in figure \ref{fig: allorderreduced 2 and 3}(b).

After some rewriting and using the color singlet properties of the external states, the sum of the contributions in eq. \eqref{eq: usoft melts} is proportional to 
\begin{align}
\langle \f O_{\bar c c}^{\mu} \rangle_{{\rm figure}\,\ref{fig: allorderreduced 2 and 3}(b)}&= \frac{1}{8\pi^2} \int_{-\infty}^{\infty} d\bar n \cdot l \, dn \cdot l\, {\rm tr}\Big ( \f I_{\bar c \perp}^{\nu}(\bar n \cdot l) \gamma_{\perp}^{\mu} \f I_c (n \cdot l) \f I_{us}(\bar n \cdot l , n \cdot l) \Big ),
\label{eq: Obarcc fig5}
\end{align}
where the collinear, anti-collinear, and ultrasoft matrix elements are
\begin{align} \notag
\f I_c(n \cdot l) &= \int d^4z \, e^{\frac{i}{2} \bar n \cdot z \, n \cdot l} \langle p' | T \chi_c(0) \left [\bar \xi_c i \overleftarrow{\s D}_{c\perp} W_c \right ](z) | p \rangle,
\\
\f I_{\bar c \perp}^{\nu}(\bar n \cdot l) &= \int d^4z \, e^{\frac{i}{2} n \cdot z \, \bar n \cdot l} \langle 0 | T \Big ( \gamma_{\perp}^{\nu} e^{iz \cdot q'} \chi_{\bar c}(z) + \left [ W_{\bar c}^{\dagger} i \s D_{\bar c\perp} \xi_{\bar c} \right ](z) \, \f T_{\bar c\bar c \perp}^{\nu} \Big ) \bar \chi_{\bar c}(0)  | 0 \rangle,
\label{eq: I functions}
\\ \notag
\f I_{us}(\bar n \cdot l , n \cdot l) &= \frac{1}{2} \intR dn \cdot z \, d\bar n \cdot z \, e^{-\frac{i}{2} (\bar n \cdot z \, n \cdot l + n \cdot z \, \bar n \cdot l)} 
\\ \notag
&\quad \times \langle 0 |  T q_{us}\left (\bar n \cdot z \frac{n}{2} \right )  \bar q_{us}\left ( n \cdot z \frac{\bar n}{2} \right ) Y_{\rm cusp}\left ( n \cdot z \frac{\bar n}{2}, \bar n \cdot z \frac{n}{2} \right )| 0 \rangle.
\end{align}
We introduced 
\begin{align}
Y_{\rm cusp}(s_1 n, s_2 \bar n) = Y_n^{\dagger}(s_1 n) Y_n(0) Y_{\bar n}^{\dagger}(0) Y_{\bar n}(s_2 \bar n),
\label{eq: Ycusp}
\end{align}
which is a Wilson line that goes from $s_2 \bar n$ to the origin along the $\bar n$-light-cone and then from the origin to $s_1 n$ along the $n$-light-cone. 
In eq. \eqref{eq: I functions} each $\f I_j$ is a matrix in Dirac space, but each is traced individually over color indices. Note that the factor $Y_n(0) Y_{\bar n}^{\dagger}(0)$, which originated from the decoupling transformation in the hard current $\f O_{\bar c c}^{\mu}$, cancels the Wilson lines segments that extend to infinity resulting in a finite length Wilson line. 

The function $\f I_{\bar c \perp}^{\nu}$ is known commonly as the radiative jet function. The standard form \cite{Liu:2020ydl} is recovered after undoing the contraction of the photon field with the external state (this also generalizes $\f I_{\bar c \perp}$ to higher orders in the QED coupling $\alpha_{\rm em} = e^2/4\pi$), i.e.
\begin{align}
\epsilon_{\perp\nu}^*(q') \f I_{\bar c \perp}^{\nu}(\bar n \cdot l) = \frac{1}{e} \int d^4z \, e^{\frac{i}{2} n \cdot z\, \bar n \cdot l} \bra{\gamma(q')} T \left [W_{\bar c}^{\dagger} \left (i \s D_{\bar c\perp} + e \s A_{\bar c}^{(\gamma)} \right ) \xi_{\bar c} \right ](z) \bar \chi_{\bar c}(0) \ket 0,
\end{align}
where $A_{\bar c}^{(\gamma)}$ is the anti-collinear photon field.
The radiative jet function was also identified in the context of DVCS in \cite{Schoenleber:2022myb} and is known to two loops \cite{Liu:2020ydl}.

After multiplying with the matching coefficient of the Sudakov form factor 
\begin{align}
C_V(-q^2) = 1 + \frac{\alpha_s C_F}{4\pi} \Big ( - \log^2\frac{-q^2}{\mu^2} + 3 \log \frac{-q^2}{\mu^2} + \frac{\pi^2}{6} - 8 \Big ) + \f O(\alpha_s^2),
\end{align}
which is known to three loops \cite{Baikov:2009bg, Gehrmann:2010ue},
one obtains the contribution to the hadronic tensor
\begin{align} \notag
T_{{\rm figure}\,\ref{fig: allorderreduced 2 and 3}(b)}^{\mu \nu} &= \frac{i}{2} C_V(-q^2) \sum_{(\Gamma, \bar \Gamma)} \intR d\bar n \cdot l \, \text{tr} \Big [ \f I_{\bar c \perp}^{\nu}(\bar n \cdot l)\gamma_{\perp}^{\mu} \frac{\Gamma}{2} \Big ] 
\\
&\quad \times \frac{1}{8\pi^2 i}\intR dn \cdot l \, \text{tr} \Big [ \f I_c(n \cdot l) \f I_{us}(\bar n \cdot l , n \cdot l) \frac{\bar \Gamma}{2} \Big ],
\label{eq: Tfig5}
\end{align}
where $(\Gamma, \bar \Gamma) \in \{ (n, \bar n), (n \gamma_5, \gamma_5 \bar n) \}$.

In \cite{Schoenleber:2022myb} it was shown that the leading term as $x \rightarrow \eta$ in the leading power coefficient function $C_{\bar \chi \chi}^{[\Gamma]}$ is given by $\f I_{\bar c \perp}^{\nu}$ multiplied by the matching coefficient of the hard Sudakov form factor, i.e.
\begin{align}
C_{\bar \chi \chi}^{[\Gamma] \mu \nu}(x) = \frac{i}{2} C_V(-q^2) \text{tr} \Big [ \f I_{\bar c \perp}^{\nu}(\bar c \cdot l)\gamma_{\perp}^{\mu} \frac{\Gamma}{2} \Big ] + \f O((\bar n \cdot l)^0).
\end{align}
It is questionable whether the remaining factors in eq. \eqref{eq: Tfig5} describing the ultrasoft and collinear modes can be interpreted as universal non-perturbative objects. This is because the ultrasoft quark, which is formally part of the hadron, can interact with the ultrasoft gluons dressing the virtual photon vertex, which is encoded in the  $Y_n(0)Y_{\bar n}^{\dagger}(0)$ factor in eq. \eqref{eq: Ycusp}. This is similar to the lack of universality of the B-meson light-cone distribution function observed in the context of QED corrections \cite{Beneke:2017vpq,Beneke:2019slt}.

Note that eq. \eqref{eq: Tfig5} presents an instance of what can be referred to as ``endpoint-refactorization'', meaning that the perturbative function factorizes in a certain kinematic limit, where a partons momentum becomes small (in this case of the internal loop momentum variable $\bar n \cdot l = (x- \eta) \bar n \cdot P$). As mentioned, in the case of DVCS, this was also identified \cite{Schoenleber:2022myb} (without the use of SCET). For recent discussion of endpoint-factorization in the context of other processes, consider for example \cite{Bell:2022ott, Feldmann:2022ixt}.

We remark that $\langle \f O_{\bar c c}^{\mu} \rangle_{{\rm figure}\,\ref{fig: allorderreduced 2 and 3}(b)}$ is zero to all orders in perturbation theory, assuming massless quarks. Indeed, note that $\langle \f O_{\bar c c}^{\mu} \rangle_{{\rm figure}\,\ref{fig: allorderreduced 2 and 3}(b)}$ is proportional to $\s n \f I_{us} \sbar n$ (note that $\f I_c = \f I_c \frac{\sbar n \s n}{4}$ and $\f I_{\bar c} = \frac{\sbar n \s n}{4} \f I_{\bar c}$). If we have exact chiral symmetry, we can write
\begin{align}
\f I_{us}(n \cdot l , \bar n \cdot l) = \f I_{usV}^{\alpha}(n \cdot l , \bar n \cdot l) \gamma_{\alpha} + \f I_{usA}^{\alpha}(n \cdot l , \bar n \cdot l) \gamma_{\alpha} \gamma_5.
\label{eq: Is dirac decomp}
\end{align}
Thus, since $\f I_{us}$ is sandwiched between $\s n$ and $\sbar n$, only the $\f I_{usV \perp}^{\alpha}$ and $\f I_{usA \perp}^{\alpha}$ can contribute. But, since there is no non-zero transverse vector to carry the $\alpha$ index, we must have $\f I_{usV \perp}^{\alpha} = \f I_{usA \perp}^{\alpha} = 0$. However, since the chiral symmetry is spontaneously broken, we can have more Dirac structures than in eq. \eqref{eq: Is dirac decomp}. For instance, the one proportional to the unit matrix in Dirac space
\begin{align}
\f I_{us \mathbb I}(\bar n \cdot l , n \cdot l) &= \frac{1}{2} \intR dn \cdot z \, d\bar n \cdot z \, e^{-\frac{i}{2} (\bar n \cdot z \, n \cdot l + n \cdot z \, \bar n \cdot l)} 
\\ \notag
&\quad \times \langle 0 |  T \bar q_{us}\left ( n \cdot z \frac{\bar n}{2} \right ) Y_{\rm cusp}\left ( n \cdot z \frac{\bar n}{2} , \bar n \cdot z \frac{n}{2}\right ) q_{us}\left (\bar n \cdot z \frac{n}{2} \right )  | 0 \rangle.
\end{align}
Since this ultrasoft matrix element has quantum numbers of the vacuum, it can receive contributions from chiral condensate. Thus it is generally non-zero with $\f I_{us\mathbb I} \sim \Lambda_{\rm QCD}$ (more precisely, the pion decay constant $4\pi f_{\pi}$) so that the parametrical size of $T^{\mu \nu}_{{\rm figure}\,\ref{fig: allorderreduced 2 and 3}(b)}$ is given by $\lambda \sim \Lambda_{\rm QCD}/Q$, i.e. the same size as the other NLP contributions (given that $\sqrt{-t} \sim \Lambda_{\rm QCD}$). 

It is important to remark that there is an utterly analogous ultrasoft contribution in the crossed ($u$) channel, which corresponds to a region where the other $n$-collinear parton becomes ultrasoft (i.e., $x + \eta \sim \lambda^2$ instead of $x - \eta \sim \lambda^2$). Since the discussion is entirely analogous, we did not consider this here.

Now, let us interpret the technical result of the contribution $\langle \f O_{\bar c c}^{\mu} \rangle_{{\rm figure}\,\ref{fig: allorderreduced 2 and 3}(b)} \sim \lambda^1$ in the context of DVCS phenomenology. First of all, we need to address the issue of the unphysicality of the ultrasoft modes, which calls into question the actual correctness of eq. \eqref{eq: Obarcc fig5} beyond perturbation theory. More correctly, one should do the analysis using the soft modes instead of ultrasoft. In practice, the soft region typically appears in conjunction with intermediate hard-collinear virtuality and the EFT setup involves two step matching where one first integrates-out hard modes and obtains SCET$_{\rm I}$ setup and then in the second step the hard-collinear modes are integrated out leading to SCET$_{\rm II}$ which contains soft and collinear modes. While this is in principle straightforward, it is more involved than we have done. 

There exists another powerful tool for the study of such ``soft'' effects in the so-called light-cone sum rules. This technique has been applied to the closely related process of $\pi^0 \rightarrow \gamma^* \gamma$, which is essentially the crossed process of DVCS, see e.g. \cite{Agaev:2010aq}. Note that the light-cone sum rule analysis is more complicated for DVCS due to the different causal structure of having both initial and final state hadrons. We leave connecting  the light-cone sum rules with SCET framework in the context of DVCS for future work.

Last but not least, it should be remarked on an important distinction based on the polarization of the incoming virtual photon $\gamma^*$. First note that the leading $\sim \lambda^0$ collinear contribution is only non-zero for transverse polarizations of $\gamma^*$. To see this, let $C_{LP}$ be a coefficient function of a leading power operator in eq. \eqref{eq: LP operators}. In addition to $\mu$ and $\nu = \perp$ it can have an additional two indices for the two-gluon operator in eq. \eqref{eq: LP operators}. Suppose that $\mu \neq \perp$. Then $C_{LP}$ must have an odd number of transverse indices. In two dimensions the available covariant tensors are $g_{\perp}^{\mu_1 \mu_2}$ and $\varepsilon_{\perp}^{\mu_1 \mu_2}$, which both have an even number of indices. Thus $C_{LP}$ must be proportional to a transverse vector. This can only be $q_{\perp}$ or $q_{\perp}'$ which are both zero. Thus, for the longitudinal polarization $\lambda^1$ is actually the leading power. 

On the other hand, we have $n_{\mu} C_V \langle \f O_{\bar c c}^{\mu} \rangle_{{\rm figure}\,\ref{fig: allorderreduced 2 and 3}(b)} = \bar n_{\mu} C_V \langle \f O_{\bar c c}^{\mu} \rangle_{{\rm figure}\,\ref{fig: allorderreduced 2 and 3}(b)} = 0$, so the ultrasoft $\sim \lambda^1$ contribution actually vanishes for longitudinal virtual photon polarizations and therefore the NLP DVCS amplitude is actually given in terms of the collinear operators in eq. \eqref{eq: NLP operators}. Note that this fact does not change if the low energy modes that mediate between the collinear and anti-collinear sectors are soft instead of ultrasoft. We have already verified this for the one-loop example in eq. \eqref{eq: soft pc}. More generally, in the soft case the virtual photon must undergo hard-collinear to hard-anti-collinear splitting, i.e. we have the same hard current $\f O_{\bar c c}^{\mu}$ but with the collinear and anti-collinear quarks fields replaced by hard-collinear and hard-anti-collinear quark fields. Hence the $\mu$ index is also projected onto its transverse component, leading to additional suppression for $\mu \neq \perp$. To show this more rigorously one must perform the matching onto SCET$_{\rm II}$ which is beyond the scope of this work.

We can say even more by considering the Lorentz structures of a coefficient function $C_{\rm NLP}$ of an NLP operator in eq. \eqref{eq: NLP operators}. Clearly $C_{\rm NLP}$ must have an odd number of indices, i.e. in addition to $\mu$ and $\nu = \perp$ either one or three additional indices for the operators involving quarks and only gluons respectively. Suppose $\mu = \perp$. Because $C_{\rm NLP}$ has an odd number of transverse indices it must depend on a transverse vector, implying that it is zero, since $q_{\perp} = q_{\perp}' = 0$. Thus, $C_{\rm NLP}$ can only be non-zero for a longitudinally polarized virtual photon. 

In conclusion we have argued for the following two statements
\begin{itemize}
    \item[$(i)$] The DVCS amplitude $T^{\mu \nu}$ for $\mu \neq \perp$ and $\nu = \perp$ is $\sim \lambda^1$ and at leading power it factorizes in terms of the collinear twist-3 operators in eq. \eqref{eq: NLP operators}. In particular there is no (ultra)soft contribution at this accuracy which implies the absense of endpoint divergences to all orders in $\alpha_s$.
    \item[$(ii)$] The DVCS amplitude $T^{\mu \nu}$ for $\mu,\nu = \perp$ is $\sim \lambda^0$ and the leading contribution factorizes in terms of the collinear operators in eq. \eqref{eq: LP operators}. The NLP $\sim \lambda^1$ contribution is given entirely by a contribution corresponding to the reduced graph in figure \ref{fig: allorderreduced 2 and 3}(b). 
\end{itemize}
We stress that these statements are only true in the BMP frame. However, the factorization is not ``broken'' by going to other frames related to the BMP frame through RPI transformations. In other words, this implies the factorization also in the other frames.

\section{Tree-level calculation}
\label{sec: calc}

\begin{figure}
\centering
\includegraphics[scale=.35]{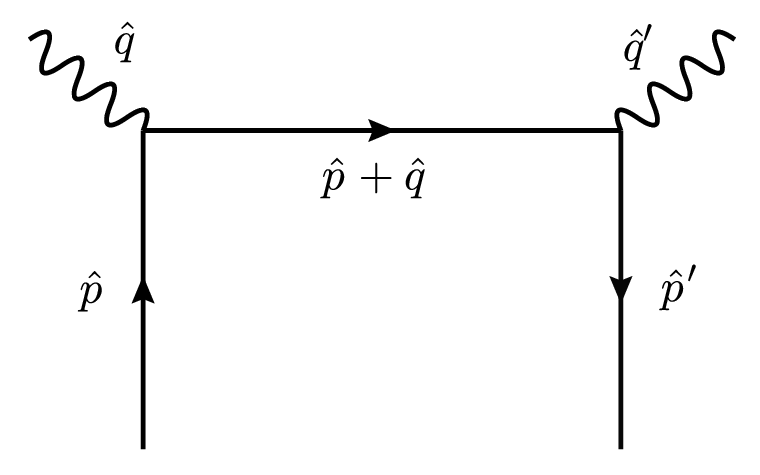} \hspace{.5cm}
\includegraphics[scale=.4]{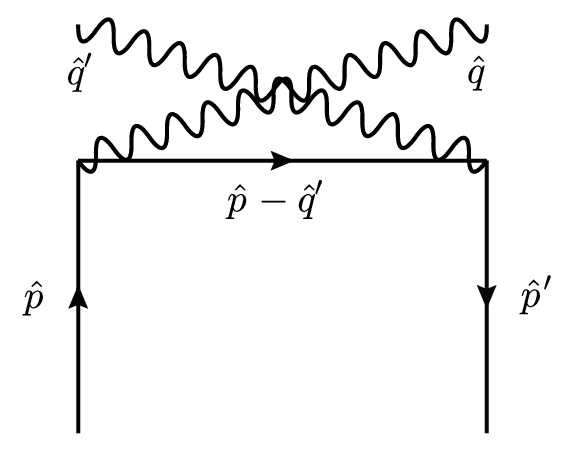}
\caption{Two-quark leading order QCD diagrams.}
\label{fig: qqmatching}
\end{figure}

To further illustrate the here-developed formalism and establish a connection with the traditional QCD approach, we present the calculation of the hard matching coefficients at leading order in $\alpha_s$.

\subsection{Two-quark external states}

We evaluate both sides of eq. \eqref{eq: factorization theorem} with a two-quark external states at $\alpha_s^0$ accuracy to obtain the coefficient functions $C_{\bar \chi \chi}^{[\Gamma]\mu \nu(0)}$ and $C_{\bar \chi \p \chi}^{[\Gamma]\mu \nu \alpha (0)}$. Let
\begin{align}
\hat p^{\mu} &= (x+ \eta) \bar n \cdot P \, \frac{n^{\mu}}{2} + P_{\perp}^{\mu}, \notag
\\
\hat p'^{\mu} &= (x - \eta) \bar n \cdot P \, \frac{n^{\mu}}{2} + P_{\perp}^{\mu},\notag
\\
\hat q^{\mu} &= - (\rho + \eta) \bar n \cdot P \frac{n^{\mu}}{2} + \frac{Q^2}{\rho \bar n \cdot P} \frac{\bar n^{\mu}}{2} ,
\\
\hat q'^{\mu} &= - (\rho - \eta) \bar n \cdot P \frac{n^{\mu}}{2} + \frac{Q^2}{\rho \bar n \cdot P} \frac{\bar n^{\mu}}{2}.\notag
\end{align} 
We put the hats on the momenta to distinguish them from the physical momenta, but we will drop this notation for the rest of the section. It is implied that we are considering parton momenta.

The tree-level SCET matrix elements are
\begin{align}
\bra{p'} \f O_{\bar \chi \chi}^{[\Gamma]}(-s/2, s  /2) \ket {p}^{(0)} &= e^{-is \bar n \cdot P} \bar u_c(p') \frac{\Gamma}{2} u_c(p), \notag
\\
\bra{p'} \f O_{\bar \chi \p \chi}^{[\Gamma]\alpha}(-s/2, s/2) \ket {p}^{(0)} &= e^{-is \bar n \cdot P} \, P_{\perp}^{\alpha} \bar u_c(p') \frac{\Gamma}{2} u_c(p),
\end{align}
where the superscript $(0)$ denotes the tree-level matrix elements.
Thus
\begin{align}
F_{\bar \chi \chi}^{[\Gamma] (0)}(y) &= \delta(x-y) \frac{\bar u_c(p') \Gamma u_c(p)}{2 \bar n \cdot P}, \notag
\\
F_{\bar \chi \p \chi}^{[\Gamma] \alpha (0) }(y) &= \delta(x-y) \frac{P_{\perp}^{\alpha}}{\bar n\cdot P} \frac{\bar u_c(p') \Gamma u_c(p)}{2 \bar n \cdot P}. 
\end{align}
Then, the right-hand side of eq. \eqref{eq: factorization theorem} reads
\begin{align} 
T_{qq}^{\mu \nu(0)} &= \sum_{\Gamma \in \{ \sbar n, \sbar n \gamma_5 \}} \Bigg \{   C_{\bar \chi \chi}^{[\Gamma]\mu \nu (0) }(x)  \frac{\bar u_c(p') \Gamma u_c(p)}{2\bar n \cdot P} + C_{\bar \chi \p \chi}^{[\Gamma] \mu \nu \alpha(0)}(x) \frac{P_{\perp \alpha}}{\bar n\cdot P} \frac{\bar u_c(p') \Gamma u_c(p)}{2 \bar n \cdot P} \Bigg \}.
\end{align}
The QCD contribution is given by the diagrams in figure \ref{fig: qqmatching}:
\begin{align} \notag
T_{qq}^{\mu \nu(0)} &= i\bar u(p') \Bigg \{  \gamma^{\nu} \frac{i}{\s p + \s q} \gamma^{\mu} + \gamma^{\mu} \frac{i}{\s p' - \s q} \gamma^{\nu} \Bigg \}  u(p)
\\ \notag
&= \frac{\bar u_c \sbar n u_c}{\bar n \cdot P} \Bigg \{ - g_{\perp}^{\mu \nu} \frac{x}{\rho^2 - x^2} 
+ \frac{P_{\perp \alpha} }{\bar n \cdot P} \Bigg ( \frac{-(\eta \rho + x^2) g_{\perp}^{\mu \alpha} \bar n^{\nu} + (\eta \rho - x^2) g_{\perp}^{\nu \alpha} \bar n^{\mu}}{(\rho^2 - x^2)(\eta^2 - x^2)} 
\\
&\qquad + \frac{(\bar n \cdot P)^2}{Q^2} \frac{\rho(\eta - \rho)(\eta \rho + x^2) g_{\perp}^{\mu \alpha} n^{\nu} + \rho (\eta + \rho) (\eta \rho - x^2) g_{\perp}^{\nu \alpha} n^{\mu}}{(\rho^2 - x^2)(\eta^2 - x^2)}\Bigg ) \Bigg \}
\\ \notag
&\quad - i\frac{\bar u_c \sbar n \gamma_5 u_c}{\bar n \cdot P} \Bigg \{ \varepsilon_{\perp}^{\mu \nu} \frac{ \rho}{\rho^2 - x^2} 
+ \frac{P_{\perp \alpha}}{\bar n \cdot P}  \Bigg (  \frac{x(\eta - \rho) \varepsilon_{\perp}^{\nu \alpha} \bar n^{\mu} + x (\eta + \rho) \varepsilon_{\perp}^{\mu \alpha} \bar n^{\nu}}{(\rho^2 - x^2)(\eta^2 - x^2)}
\\ \notag
&\qquad+ \frac{(\bar n \cdot P)^2}{Q^2} \frac{x\rho (\eta^2 - \rho^2) ( \varepsilon_{\perp}^{\nu \alpha} n^{\mu} - \varepsilon_{\perp}^{\mu \alpha} n^{\nu})}{(\rho^2 - x^2)(\eta^2 - x^2)} \Bigg ) \Bigg \} + \f O(\lambda^2).
\end{align}
We  now read off the coefficients functions:
\begin{align} \notag
C_{\bar \chi \chi}^{[\sbar n]\mu \nu(0)}(x) &= - 2 \frac{xg_{\perp}^{\mu \nu}}{\rho^2 - x^2}, 
\\ \notag
C_{\bar \chi \chi}^{[\sbar n \gamma_5]\mu \nu(0)}(x) &= -2i  \frac{\rho\varepsilon_{\perp}^{\mu \nu}}{\rho^2 - x^2}, 
\\ \label{eq: CF WW}
C_{\bar \chi \p \chi}^{[\sbar n]\mu \nu \alpha(0)}(x) &= -2 \frac{(x^2 + \eta \rho) g_{\perp}^{\mu \alpha} \bar n^{\nu} + (x^2 - \eta \rho) g_{\perp}^{\nu \alpha} \bar n^{\mu}}{(\rho^2 - x^2)(\eta^2 - x^2)}
\\ \notag
&\qquad - 2\kappa \frac{\rho(\rho - \eta)(x^2 + \eta \rho) g_{\perp}^{\mu \alpha} n^{\nu} + \rho (\rho + \eta) (x^2 - \eta \rho) g_{\perp}^{\nu \alpha} n^{\mu}}{(\rho^2 - x^2)(\eta^2 - x^2)},
\\ \notag
C_{\bar \chi \p \chi}^{[\sbar n \gamma_5]\mu \nu \alpha(0)}(x) &= -2i \frac{x (\rho + \eta) \varepsilon_{\perp}^{\mu \alpha} \bar n^{\nu} - x(\rho - \eta) \varepsilon_{\perp}^{\nu \alpha} \bar n^{\mu}}{(\rho^2 - x^2)(\eta^2 - x^2)} -2i \kappa \frac{x\rho (\rho^2 - \eta^2) ( \varepsilon_{\perp}^{\mu \alpha} n^{\nu} - \varepsilon_{\perp}^{\nu \alpha} n^{\mu})}{(\rho^2 - x^2)(\eta^2 - x^2)},
\end{align}
where $\kappa = \frac{(\bar n \cdot P)^2}{Q^2}$.

We emphasize that in the SCET framework, the Feynman pole prescription $i0$ must be ``inherited'' by the on-shell QCD amplitude. Throughout this section, we have dropped the $i0$.
To restore it, note that 
\begin{align}
(p+q)^2 + i0 &= - (\rho - x - \text{sgn}(Q^2/\rho)\, i0) Q^2/\rho, \notag
\\
(p - q')^2 + i0 &= - (\rho + x - \text{sgn}(Q^2/\rho)\, i0) Q^2/\rho.
\end{align}
For DVCS, we have $\rho = \eta > 0$ and $Q^2 > 0$ so that the pole prescription can be recovered by taking $\rho \rightarrow \rho - i0$. On the other hand, the $\frac{1}{\eta \pm x}$ denominators arose from eq. \eqref{eq: u to un}, so the prescription for this pole is a matter of choice, and the final result should not depend on it.

The DVCS coefficient functions are obtained by setting $\eta = \rho$ and $\nu = \perp$.
\begin{align} \notag
C_{\bar \chi \chi}^{[\sbar n]\mu \nu(0)}(x) &= - \frac{2xg_{\perp}^{\mu \nu}}{(\eta - x - i0)(\eta + x - i0)}, 
\\ \notag
C_{\bar \chi \chi}^{[\sbar n \gamma_5]\mu \nu(0)}(x) &= - \frac{2\eta i\varepsilon_{\perp}^{\mu \nu}}{(\eta - x - i0)(\eta + x - i0)}, 
\\ \label{eq: CF WW DVCS}
C_{\bar \chi \p \chi}^{[\sbar n]\mu \nu \alpha(0)}(x) &= \frac{2g_{\perp}^{\nu \alpha} \bar n^{\mu} + 4\kappa\eta^2 g_{\perp}^{\nu \alpha} n^{\mu}}{(\eta - x - i0)(\eta + x - i0)},
\\ \notag
C_{\bar \chi \p \chi}^{[\sbar n \gamma_5]\mu \nu \alpha(0)}(x) &= 0,
\end{align}
where we have explicitly included the $i0$ prescription.
Interestingly, the $\sim \lambda^1$ axial-vector contribution vanishes. It is not clear whether this continues at higher orders. There are only simple poles at $x = \pm \eta$. Thus, if $F_{\bar \chi \p \chi}^{[\sbar n] \alpha}$ is continuous at $x \pm \eta$, the convolution integral is well-defined. It is well-known that certain twist-3 GPDs are in fact discontinuous at certain points. However, it is also well-known \cite{Kivel:2000cn} that the convolution integrals are convergent nevertheless. We will see later, in section \ref{sec: old approach}, that the coefficient functions in eq. \eqref{eq: CF WW DVCS} agree with the result from \cite{Kivel:2000cn}.

\subsection{Quark-gluon-quark external states}
\begin{figure}
\centering
\includegraphics[scale=.2]{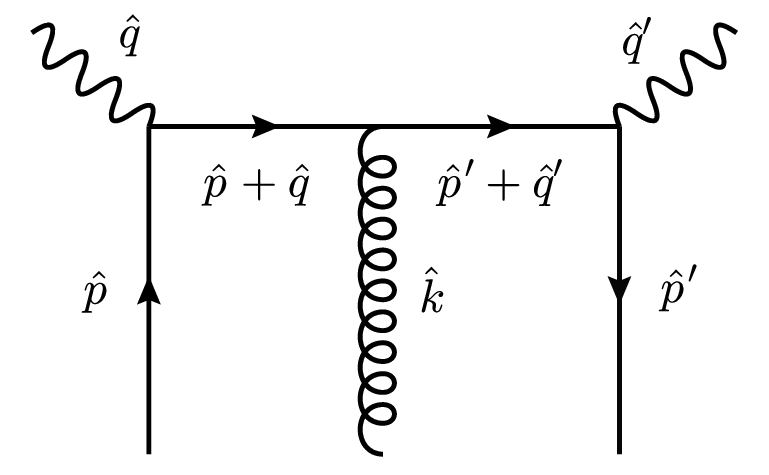} \qquad
\includegraphics[scale=.2]{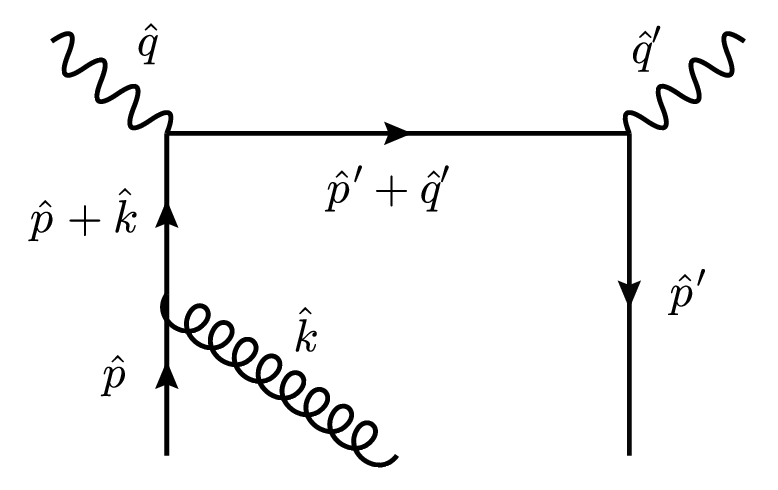} \qquad
\includegraphics[scale=.16]{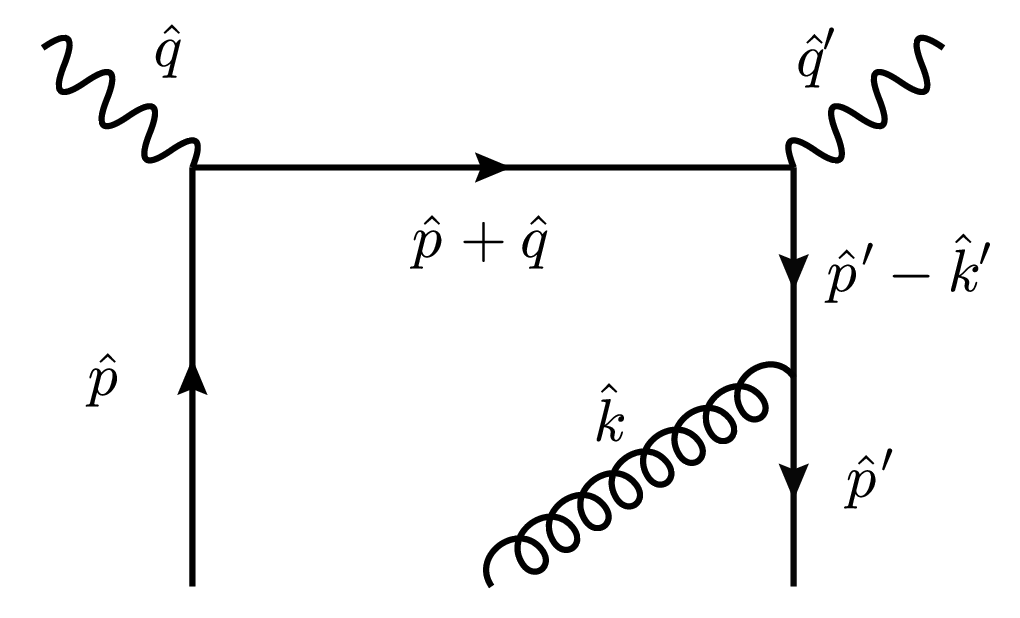}
\\
\hspace{1cm} (a) \hspace{4.3cm} (b) \hspace{4.3cm}(c) \hspace{1cm}
\caption{Quark-gluon-quark QCD diagrams. Diagrams with crossed photon legs also contribute but are not shown.}
\label{fig: qgq dias}
\end{figure}
\begin{figure}
\centering
\includegraphics[scale=.2]{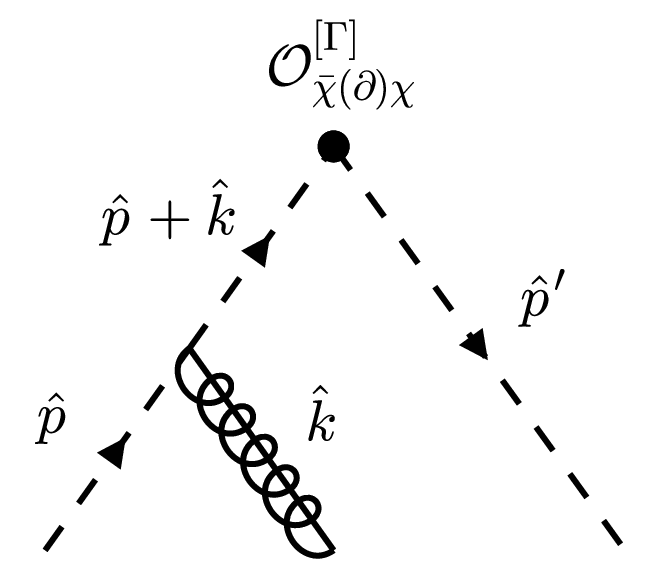} \qquad \includegraphics[scale=.24]{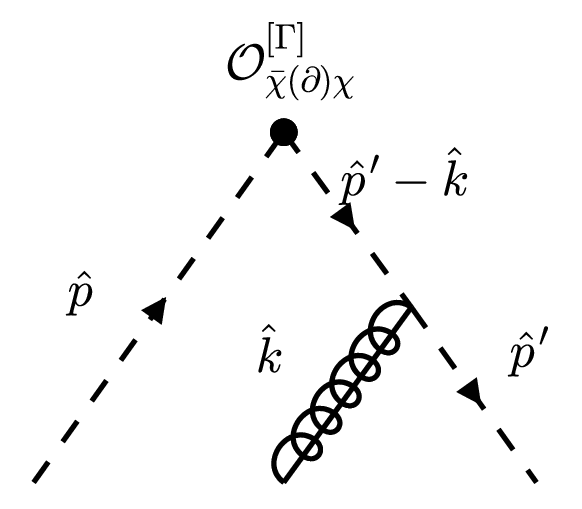}
\\
\hspace{4.8cm} (a) \hspace{3.8cm} (b) \hspace{5cm}
\caption{SCET diagrams contributing to the matching with quark-gluon-quark external states. At the top vertex we can have insertions of $\f O_{\bar \chi \chi}^{[\Gamma]}$ or $\f O_{\bar \chi \p \chi}^{[\Gamma]\mu}$. Dashed propagators are $n$-collinear quarks, and the gluon propagator with lines through them denotes $n$-collinear gluons.}
\label{fig: qgq SCET dias}
\end{figure}
Let us choose external parton states with momenta
\begin{align} \notag
\hat p^{\mu} &= (x_1 -x_2+ \eta) \bar n \cdot P \, \frac{n^{\mu}}{2} + P_{\perp}^{\mu}\,,
\\ \notag
\hat p'^{\mu} &= (x_1 + x_2- \eta) \bar n \cdot P \, \frac{n^{\mu}}{2} + P_{\perp}^{\mu}\,,
\\
\hat k^{\mu} &= 2x_2 \bar n \cdot P \frac{n^{\mu}}{2}\,,
\label{eq: qgq momenta}
\\ \notag
\hat q^{\mu} &= - (\rho + \eta) \bar n \cdot P \frac{n^{\mu}}{2} + \frac{Q^2}{\rho \bar n \cdot P} \frac{\bar n^{\mu}}{2} \,,
\\ \notag
\hat q'^{\mu} &= - (\rho - \eta) \bar n \cdot P \frac{n^{\mu}}{2} + \frac{Q^2}{\rho \bar n \cdot P} \frac{\bar n^{\mu}}{2}\,.
\end{align}
Again, we drop the hats for the rest of this subsection, tacitly implying that we are considering partonic momenta. Here $\hat k$ is the ingoing momentum of the $n$-collinear gluon. The QCD diagrams are shown in figure \ref{fig: qgq dias}. We have to keep $P_{\perp}$ for the quarks to regulate the $n$-collinear propagators of the diagrams (b) and (c) in figure \ref{fig: qgq dias}. In the end, the limit $P_{\perp}^{\mu} \rightarrow 0$ should be taken since the additional gluon building block $\f A_{c\perp}$ already presents a power of $\lambda$. 
The singularity as $P_{\perp}^{\mu} \rightarrow 0$ in the diagrams (b) and (c) in figure \ref{fig: qgq dias} will be canceled by the SCET diagrams with the insertion of the leading power operator $\f O_{\bar \chi \chi}^{[\Gamma]}$ in figure \ref{fig: qgq SCET dias}. On the other hand, the insertion of $\f O_{\bar \chi \p \chi}^{[\Gamma]\alpha}$ will give a finite contribution (as $P_{\perp}^{\mu} \rightarrow 0$) to the matching. As for the polarization vector of the external gluon, we can choose it to be the unit-vector in $P_{\perp}$ direction, that is
\begin{align}
\epsilon^{\mu} = - \frac{P_{\perp}^{\mu}}{\epsilon \cdot P_{\perp}}.
\end{align}
Let us start with the SCET part of the matching. We have
\begin{align}
&\bra {p'} \f O_{\bar \chi \f A \chi}^{[\Gamma] \mu}(-s_1 /2, s_2/2, s_1/2) \ket {p,k}^{(0)} = g \epsilon_{\perp}^{\mu} \bar u_c \frac{\Gamma}{2} u_c e^{- is_1x_1 \bar n \cdot P - is_2x_2 \bar n \cdot P }
\end{align}
and therefore
\begin{align}
F_{\bar \chi \f A \chi}^{[\Gamma]\mu(0)}(y_1,y_2) = g \epsilon_{\perp}^{\mu} \frac{\bar u_c \Gamma u_c}{2\bar n \cdot P} \delta(y_1 - x_1) \delta(y_2 - x_2).
\end{align}
The contribution from the diagrams in figure \ref{fig: qgq SCET dias} reads
\begin{align} \notag
\langle \f O_{\bar \chi \chi}^{[\Gamma]} \rangle_{\text{fig}\,\ref{fig: qgq SCET dias}}&= \bra{p'} T\{  \f O_{\bar \chi \chi}^{[\Gamma]}(-s_1/2,s_1/2)
\\ 
&\quad  \left [ i \int d^4x\, \bar \xi \Big ( g \s A_{c\perp} \frac{1}{i\bar n \cdot \p} i\s \p_{\perp} + i\s \p_{\perp} \frac{1}{i\bar n \cdot \p} g \s A_{c\perp} \Big )\frac{\sbar n}{2} \xi \right ] \}\ket {p,k}^{(0)}
\\ \notag
&= \frac{g\bar u_c \Gamma u_c}{2\epsilon \cdot P_{\perp}} \Bigg \{ e^{-is_1(x_1+x_2) \bar n \cdot P}\left ( 1 + \frac{\bar n \cdot (p+k)}{\bar n \cdot p} \right ) + e^{-is_1(x_1-x_2) \bar n \cdot P}\left ( 1 + \frac{\bar n \cdot (p'-k)}{\bar n \cdot p'} \right ) \Bigg \}
\end{align}
and
\begin{align}
\langle \f O_{\bar \chi \p \chi}^{[\Gamma] \alpha}\rangle_{\text{fig}\,\ref{fig: qgq SCET dias}} &= P_{\perp}^{\alpha} \langle \f O_{\bar \chi \chi}^{[\Gamma]}\rangle_{\text{fig}\,\ref{fig: qgq SCET dias}}
= - \epsilon \cdot P_{\perp} \epsilon^{\alpha} \langle \f O_{\bar \chi \chi}^{[\Gamma]}\rangle_{\text{fig}\,\ref{fig: qgq SCET dias}}.
\end{align}
The contribution to the matching is obtained by convoluting with the coefficient functions:
\begin{align} \notag
(T_{qgq}^{\mu \nu})_{\text{fig}\, \ref{fig: qgq SCET dias}} &= \sum_{\Gamma} \intR \frac{ds_1}{2\pi} \hat C_{\bar \chi \chi}^{[\Gamma]\mu \nu(0)}(s_1)  \langle \f O_{\bar \chi \chi}^{[\Gamma]} \rangle_{\text{fig}\,\ref{fig: qgq SCET dias}} + \sum_{\Gamma} \intR \frac{ds_1}{2\pi} \hat C_{\bar \chi \p \chi}^{[\Gamma]\mu \nu \alpha(0)}(s_1)  \langle \f O_{\bar \chi \p \chi \alpha}^{[\Gamma]} \rangle_{\text{fig}\,\ref{fig: qgq SCET dias}}
\\ \notag
&= \sum_{\Gamma} \frac{g\bar u_c \Gamma u_c}{2\bar n \cdot P} \frac{1}{\epsilon \cdot P_{\perp}} \Bigg \{ C_{\bar \chi \chi}^{[\Gamma]\mu \nu(0)}(x_1+x_2)\left ( 1 + \frac{\bar n \cdot (p+k)}{\bar n \cdot p} \right ) 
\\
&\quad \quad  +C_{\bar \chi \chi}^{[\Gamma]\mu \nu(0)}(x_1-x_2)\left ( 1 + \frac{\bar n \cdot (p'-k)}{\bar n \cdot p'} \right ) \Bigg \}
\\ \notag
&\quad - \sum_{\Gamma} \frac{g\bar u_c \Gamma u_c}{2\bar n \cdot P} \epsilon_{\alpha} \Bigg \{ C_{\bar \chi \p \chi}^{[\Gamma]\mu \nu \alpha(0)}(x_1+x_2)\left ( 1 + \frac{\bar n \cdot (p+k)}{\bar n \cdot p} \right )
\\ \notag
&\quad \quad  +C_{\bar \chi \p \chi}^{[\Gamma]\mu \nu \alpha(0)}(x_1-x_2)\left ( 1 + \frac{\bar n \cdot (p'-k)}{\bar n \cdot p'} \right ) \Bigg \}.
\end{align}
The contributions from the QCD diagrams are given by
\begin{align}
(T_{qgq}^{\mu \nu})_{\text{fig}\,\ref{fig: qgq dias}(a)} &= i\bar u_c  \gamma^{\nu} \frac{i}{\s p' + \s q'} (ig \s \epsilon_{\perp}) \frac{i}{\s p + \s q} \gamma^{\mu} u_c + i\bar u_c  \gamma^{\mu} \frac{i}{\s p' - \s q} (ig \s \epsilon_{\perp}) \frac{i}{\s p - \s q'} \gamma^{\nu} u_c
\end{align}
and
\begin{align}
(T_{qgq}^{\mu\nu})_{{\rm fig}\, \ref{fig: qgq dias}(b)+(c)} &= g \bar u(p') \Bigg \{ \s \epsilon_{\perp} \frac{1}{\s p' - \s k} \left (  \gamma^{\nu} \frac{1}{\s p + \s q} \gamma^{\mu} + \gamma^{\mu} \frac{1}{\s p - \s q'} \gamma^{\nu}  \right ) \notag
\\
&\quad + \left (  \gamma^{\nu} \frac{1}{\s p' + \s q'} \gamma^{\mu}  + \gamma^{\mu} \frac{1}{\s p' - \s q} \gamma^{\nu} \right ) \frac{1}{\s p + \s k} \s \epsilon_{\perp} \Bigg \} u(p).
\end{align}
To summarize, the matching equation reads
\begin{align}
\sum_{\Gamma} g \epsilon_{\perp \alpha} \frac{\bar u_c \Gamma u_c}{2\bar n \cdot P} C_{\bar \chi \f A \chi}^{[\Gamma] \mu \nu \alpha}(x_1,x_2) =  (T_{qgq}^{\mu \nu})_{\text{fig}\,\ref{fig: qgq dias}} - (T_{qgq}^{\mu \nu})_{\text{fig}\,\ref{fig: qgq SCET dias}}.
\end{align}
A straightforward calculation gives
\begin{align} \notag
C_{\bar \chi A \chi}^{[\sbar n]\mu \nu \alpha(0)} &= 2\bar n^{\mu} g_{\perp}^{\nu \alpha} \frac{x_1^2-\left(\eta +x_2\right) \left(\rho
   -x_2\right)}{\left(-\eta +x_1-x_2\right) \left(\eta
   +x_1+x_2\right) \left(\rho -x_1-x_2\right) \left(\rho
   +x_1-x_2\right)}
\\ 
&\quad + 2n^{\mu} g_{\perp}^{\nu \alpha} \frac{\kappa  \rho  (\eta +\rho ) \left(x_1^2-\left(\eta
   +x_2\right) \left(\rho
   -x_2\right)\right)}{\left(-\eta +x_1-x_2\right)
   \left(\eta +x_1+x_2\right) \left(\rho -x_1-x_2\right)
   \left(\rho +x_1-x_2\right)} 
\\ \notag
&\quad + (\mu \leftrightarrow \nu, \rho \rightarrow- \rho),
\\ \notag
C_{\bar \chi A \chi}^{[\sbar n \gamma_5] \mu \nu \alpha(0)}&= 2\bar n^{\mu} i\varepsilon_{\perp}^{\nu \alpha} \frac{x_1 \left(\eta -\rho +2 x_2\right)}{\left(\eta
   -x_1+x_2\right) \left(\eta +x_1+x_2\right) \left(\rho
   +x_1-x_2\right) \left(-\rho +x_1+x_2\right)}
\\ 
&\quad + 2n^{\mu} i\varepsilon_{\perp}^{\nu \alpha} \frac{\kappa  \rho  x_1 (\eta +\rho ) \left(\eta -\rho
   +2 x_2\right)}{\left(-\eta +x_1-x_2\right) \left(\eta
   +x_1+x_2\right) \left(\rho -x_1-x_2\right) \left(\rho
   +x_1-x_2\right)} 
\\ \notag
&\quad + (\mu \leftrightarrow \nu, \rho \rightarrow - \rho).
\end{align}
Interestingly, we have
\begin{align}
C_{\bar \chi \p \chi}^{[\Gamma]\mu \nu\alpha (0)}(x) =  C_{\bar \chi \f A \chi}^{[\Gamma]\mu \nu\alpha (0)}(x,0).
\end{align}
This implies that, at the tree level, the complete NLP contribution, apart from pure gluon contributions, can be written as
\begin{align} \notag
T_{\rm NLP}^{\mu \nu(0)} &= \sum_{\Gamma} \intR \frac{ds_1 ds_2}{(2\pi)^2} \, \hat C_{\bar \chi \f A \chi}^{[\Gamma]\mu \nu \alpha (0)}(s_1,s_2) 
\\
&\quad \times \bra{p'} \bar \chi_c(-s_1\bar n/2) \Big (i \overleftrightarrow{\p}_{\perp}^{\mu} + \f A_{c \perp}(s_2 \bar n/2) \Big ) \frac{\Gamma}{2} \chi(s_1 \bar n/2) \ket p,
\label{eq: treelevel result is simple}
\end{align}
where $\Gamma \in \{ \sbar n , \sbar n \gamma_5 \}$ and
\begin{align}
\hat C_{\bar \chi \f A \chi}^{[\Gamma]\mu \nu \alpha}(s_1,s_2) = \intR dx_1 dx_2 \, e^{ix_1 s_1 \bar n \cdot P + ix_2 s_2 \bar n \cdot P} C_{\bar \chi \f A \chi}^{[\Gamma] \mu \nu \alpha}(x_1,x_2).
\end{align}
As it stands, there is no compelling reason to believe that this continues beyond tree level.

\section{Connection with the traditional approach}
\label{sec: old approach}
In this section, we use light-cone $\bar n \cdot A = 0$ gauge for simplicity, though working in a general gauge, while being more tedious, is a straightforward generalization.
In the traditional approach, one calculates the diagrams in figure \ref{fig: qqmatching} and \ref{fig: qgq dias}(a) in terms of the QCD correlators
\begin{align} \notag
F^{\mu}(x) &= \intR \frac{ds}{2\pi} e^{ixs\bar n \cdot P} \bra{p'} \bar \psi(-s\bar n/2) \frac{\gamma^{\mu}}{2} \psi(s\bar n/2) \ket p, 
\\ \notag
\w F^{\mu}(x) &= \intR \frac{ds}{2\pi} e^{ixs\bar n \cdot P} \bra{p'} \bar \psi(-s\bar n/2) \frac{\gamma^{\mu}}{2} \gamma_5 \psi(s\bar n/2) \ket p, 
\\  \notag
K_{\perp}^{\mu}(x) &= \frac{1}{\bar n \cdot P} \intR \frac{ds}{2\pi} e^{ixs \bar n \cdot P} \bra{p'} \bar \psi(-s\bar n/2) i \overleftrightarrow{\p}_{\perp}^{\mu} \frac{\sbar n}{2} \psi(s\bar n/2) \ket p,
\\
\w K_{\perp}^{\mu}(x) &= \frac{1}{\bar n \cdot P} \intR \frac{ds}{2\pi} e^{ixs \bar n \cdot P} \bra{p'} \bar \psi(-s\bar n/2) i \overleftrightarrow{\p}_{\perp}^{\mu} \frac{\sbar n}{2} \gamma_5 \psi(s\bar n/2) \ket p,
\label{eq: QCD correlators}
\\ \notag
S_{\perp}^{\mu}(x_1,x_2) &= \intR \frac{ds_1}{2\pi} \frac{ds_2}{2\pi} e^{i s_1 x_1 \bar n \cdot P + i s_2 x_2 \bar n \cdot P} \bra{p'} \bar \psi(-s_1\bar n/2) g A_{\perp}^{\mu}(s_2 \bar n/2) \frac{\sbar n}{2} \psi(s_1\bar n/2) \ket p,
\\ \notag
\w S_{\perp}^{\mu}(x_1,x_2) &= \intR \frac{ds_1}{2\pi} \frac{ds_2}{2\pi} e^{i s_1 x_1 \bar n \cdot P + i s_2 x_2 \bar n \cdot P} \bra{p'} \bar \psi(-s_1\bar n/2) g A_{\perp}^{\mu}(s_2 \bar n/2) \frac{\sbar n}{2} \gamma_5 \psi(s_1\bar n/2) \ket p.
\end{align}
In a general gauge, the definitions need to be modified by inserting Wilson lines, which ensure gauge invariance.

Note that the diagrams in figure \ref{fig: qgq dias} (b) and (c) are explicitly not included in the standard approach since they are one-particle reducible in the external parton legs and, therefore, do not contribute to the hard scattering amplitude. This seemingly breaks the electromagnetic Ward identity since we do not sum up all possible insertions of the photons. However, the missing terms are in some sense ``hidden'' in the $F_{\perp}^{\mu}$ and $\w F_{\perp}^{\mu}$ components \cite{Belitsky:2005qn}. The electromagnetic gauge invariance is restored after eliminating them using the equations of motion. In SCET, this step is performed at the Lagrangian level by integrating out the subleading spinor components of the quark field, which amounts to solving the equations of motion for those components so that it presents a more systematic approach to factorizing the amplitude at NLP. 

To obtain the known result from the ${\rm SCET}$ result, note that a pure collinear SCET Lagrangian after ultrasoft decoupling transformation is equivalent to the QCD Lagrangian. Therefore, the correlators in eq. \eqref{eq: QCD correlators} correspond  to the functions $F_{\ell}$ introduced in section \ref{sec: factorization}, upon replacing $\psi \leftrightarrow \xi_c$ and $A^{\mu} \leftrightarrow A_c^{\mu}$.


A minor subtelty is presented by the transverse components $F_{\perp}^{\mu}$. In ${\rm SCET}$ one has integrated out the subleading spinor components so that $\bar \xi_c \gamma_{\perp}^{\mu} \xi_c = 0$. However, since the pure collinear Lagrangian is equivalent to a copy of the QCD Lagrangian, we can formally reintroduce the subleading spinor components and obtain direct correspondence. Since an $\sbar n$ between parton fields projects them onto the leading spinor components anyway, we can identify
\begin{align} \notag
&\bar n \cdot F \longleftrightarrow F_{\bar \chi \chi}^{[\sbar n]}, \qquad \bar n \cdot \w F \longleftrightarrow F_{\bar \chi \chi}^{[\sbar n \gamma_5]},
\\
&K_{\perp}^{\mu} \longleftrightarrow F_{\bar \chi \p \chi}^{[\sbar n]\mu}, \qquad \w K_{\perp}^{\mu} \longleftrightarrow F_{\bar \chi \p \chi}^{[\sbar n \gamma_5]\mu},
\\ \notag
&S_{\perp}^{\mu} \longleftrightarrow F_{\bar \chi \f A \chi}^{[\sbar n]\mu}, \qquad \w S_{\perp}^{\mu} \longleftrightarrow F_{\bar \chi \f A \chi}^{[\sbar n \gamma_5]\mu}.
\end{align}
The functions $F_{\perp}^{\mu}$ and $\w F_{\perp}^{\mu}$ are related to $K_{\perp}^{\mu}, \w K_{\perp}^{\mu}, S_{\perp}^{\mu}, \w S_{\perp}^{\mu}$ by the QCD equations of motion. In fact, one can readily show that\footnote{Apply $\gamma^{\mu} \gamma^{\nu} \gamma^{\alpha} = g^{\mu \nu} \gamma^{\alpha} + \gamma^{\nu \alpha} \gamma^{\mu} - g^{\mu \alpha} \gamma^{\nu}+ i \varepsilon^{\mu \nu \alpha \beta} \gamma_{\beta} \gamma_5$ to $\sbar n \gamma_{\perp}^{\mu} \s D \psi = 0$ and multiply from the left by $\bar \psi$ for eq. \eqref{eq: eliminate K} or $\bar \psi \gamma_5$ for eq. \eqref{eq: eliminate wK}.}
\begin{align} \notag
&K_{\perp}^{\mu}(x) - x F_{\perp}^{\mu}(x) + i \varepsilon_{\perp}^{\mu \nu} \eta \w F_{\perp \nu}(x)
\\ \label{eq: eliminate K}
&= -\frac{1}{2} \intR dx_1 dx_2 \Bigg \{ \Big [ \delta(x-x_1+x_2) + \delta(x-x_1 - x_2) \Big ] S_{\perp}^{\mu}(x_1,x_2) 
\\ \notag
&\quad + \Big [ \delta(x - x_1 + x_2) - \delta(x-x_1 -x_2) \Big ] i \varepsilon_{\perp}^{\mu \nu} \w S_{\perp\nu}(x_1,x_2) \Bigg \}
\end{align}
and
\begin{align} \notag
&\w K_{\perp}^{\mu}(x) - x \w F_{\perp}^{\mu}(x) + i \varepsilon_{\perp}^{\mu \nu} \eta F_{\perp \nu}(x)
\\  \label{eq: eliminate wK}
&= - \frac{1}{2} \intR dx_1dx_2 \Bigg \{ \Big [ \delta(x-x_1+x_2) + \delta(x-x_1 - x_2) \Big ] \w S_{\perp}^{\mu}(x_1,x_2) 
\\ \notag
&\quad + \Big [ \delta(x - x_1 + x_2) - \delta(x-x_1 -x_2) \Big ] i \varepsilon_{\perp}^{\mu \nu} S_{\perp \nu}(x_1,x_2) \Bigg \}.
\end{align}
To obtain the standard result we can eliminate $K_{\perp}^{\mu}$ and $\w K_{\perp}^{\mu}$, by the virtue of eqs. \eqref{eq: eliminate K} and \eqref{eq: eliminate wK}, in the factorization formula eq. \eqref{eq: factorization theorem}. Hence
\begin{align} \notag
T^{\mu \nu(0)} &= \intR dx\, \Big \{ C_{ \bar \chi\chi}^{[\sbar n] \mu \nu(0)} \bar n\cdot F + C_{\bar \chi \chi}^{[\sbar n \gamma_5]\mu \nu(0)} \bar n \cdot \w F + C_{\bar \chi \p \chi}^{[\sbar n]\mu \nu\alpha(0)} K_{\perp \alpha} + C_{\bar \chi \p \chi}^{[\sbar n \gamma_5]\mu \nu \alpha(0)} \w K_{\perp \alpha}
\\ \notag
&\quad + C_{\bar \chi \f A \chi}^{[\sbar n] \mu \nu \alpha(0)} S_{\perp \alpha} + C_{\bar \chi \f A \chi}^{[\sbar n \gamma_5] \mu \nu \alpha(0)} \w S_{\perp \alpha} \Big \}
\\
&= \intR dx\, C_-  \Big \{ - g_{\perp}^{\mu \nu} \bar n \cdot F + F_{\perp}^{\mu} (\bar n^{\nu} +  \rho (\rho - \eta) \kappa n^{\nu}) + F_{\perp}^{\nu} (\bar n^{\mu} +  \rho (\rho + \eta) \kappa n^{\mu}) \Big \} 
\label{eq: old result}
\\ \notag
&\quad + \intR dx\, C_+ \Big \{ -i\varepsilon_{\perp}^{\mu \nu} \bar n \cdot \w F + i \varepsilon_{\perp}^{\mu \alpha} \w F_{\alpha}(\bar n^{\nu} + \kappa \rho (\rho - \eta)n^{\nu}) - i\varepsilon_{\perp}^{\nu \alpha} \w F_{\alpha} (\bar n^{\mu} + \kappa \rho (\rho + \eta) n^{\mu}) \Big \},
\end{align}
where
\begin{align}
C_{\pm}(x) = \frac{1}{\rho - x} \pm \frac{1}{\rho + x}.
\end{align}
Eq. \eqref{eq: old result} agrees with the known results in \cite{Kivel:2000cn}, considering that we are working the BMP frame, where $\Delta_{\perp} = 0$. The results for the other frames can readily be recovered by redefining the light-cone vectors, which is discussed in the following section \ref{sec: RPI}.

Interestingly, the quark-gluon-quark contribution cancels entirely at the tree level. Of course, this observation is far from new, see e.g. \cite{Anikin:2000em, Penttinen:2000dg, Belitsky:2000vx, Radyushkin:2000jy}.
It is equivalent to the statement in eq. \eqref{eq: treelevel result is simple}. As mentioned, we do not see why this should continue at higher orders.

\section{RPI and transformation to other frames}
\label{sec: RPI}

Throughout this section, we shall drop terms of $\f O(\lambda^2)$. 
The central statement of reparametrization invariance is that observables -- in this case, we apply this to the hadronic tensor $T^{\mu \nu}$ -- are invariant under certain redefinitions of the light-cone vectors $n$ and $\bar n$, that preserve the scaling of the external momenta. See, for example, \cite{Marcantonini:2008qn} for a characterization of such transformations.

A general RPI transformation takes one from $(n, \bar n)$ light-cone basis to another light-cone system $(w, \bar w)$. A generic vector $v^{\mu}$ can then be written in both coordinates as
\begin{align}
v^{\mu} = n \cdot v \frac{\bar n^{\mu}}{2} + \bar n \cdot v \frac{n^{\mu}}{2} + v_{\perp}^{\mu} = w \cdot v \frac{\bar w^{\mu}}{2} + \bar w \cdot v \frac{w^{\mu}}{2} + v^{\perp \mu}.
\end{align}
Naturally, we demand $w^2 = \bar w^2 = w \cdot \bar w - 2 = 0, \, w \cdot v^{\perp} = \bar w \cdot v^{\perp} = 0$.
Furthermore, we define
\begin{align}
g^{\perp \mu \nu} = g^{\mu \nu} - \frac{1}{2} w^{\mu} \bar w^{\nu} - \frac{1}{2} w^{\nu} \bar w^{\mu}, \qquad \varepsilon^{\perp \mu \nu} = \varepsilon^{\mu \nu \alpha \beta} \frac{1}{2} w_{\alpha} \bar w_{\beta}.
\end{align}
We start by converting our result from the BMP to the KP frame. The transformation that takes one from the BMP $(n, \bar n)$ frame to the KP $(w, \bar w)$ frame is given by
\begin{align} \notag
&n^{\mu} = w^{\mu} + \delta_1^{\mu}, \qquad \bar n^{\mu} = \bar w^{\mu} + \delta_2^{\mu}, \qquad 
\\ \label{eq: n to w}
&v^{\perp \mu} = v_{\perp}^{\mu} + \delta_1 \cdot v \, \frac{\bar n^{\mu}}{2} + \delta_2 \cdot v \frac{n^{\mu}}{2} + \bar n \cdot v \frac{\delta_1^{\mu}}{2} + n \cdot v \frac{\delta_2^{\mu}}{2},
\end{align}
where
\begin{align}
\delta_1^{\mu} = \frac{2\Delta^{\perp\mu}}{\bar n \cdot \Delta} = - \frac{2P_{\perp}^{\mu}}{\bar n \cdot P}, \qquad \delta_2^{\mu} = - \frac{2\Delta^{\perp \mu}  \bar n \cdot q}{\bar n \cdot \Delta \, n \cdot q' }.
\end{align}
Since $\delta_1 \sim \delta_2 \sim \lambda$, this does not change the scaling of external momenta and is an allowed RPI transformation. 
One can easily compute
\begin{align}
g_{\perp}^{\mu \nu} &= g^{\perp \mu \nu} + \frac{1}{2 \eta \bar n \cdot P} (\Delta^{\perp \mu} \bar w^{\nu} + \Delta^{\perp \nu} \bar w^{\mu} ) + \frac{\rho(\rho + \eta)}{2\eta \bar n \cdot P}  \frac{(\bar n\cdot P)^2}{Q^2} ( w^{\mu} \Delta^{\perp \nu} + w^{\nu} \Delta^{\perp \mu}),
\\ \notag
\varepsilon_{\perp}^{\mu \nu} &= \varepsilon^{\perp \mu \nu} - \frac{\bar w^{\mu} \varepsilon^{\perp \nu \alpha}  \Delta^{\perp}_{ \alpha}}{2 \eta \bar n \cdot P} + \frac{\bar w^{\nu}\varepsilon^{\perp\mu \alpha}  \Delta^{\perp}_{ \alpha}}{2\eta \bar n \cdot P} 
\\
&\quad - \rho (\rho + \eta) \frac{(\bar n \cdot P)^2}{Q^2} \frac{w^{\mu} \varepsilon^{\perp \nu \alpha}\Delta^{\perp}_{\alpha}}{2\eta \bar n \cdot P} + \rho (\rho + \eta) \frac{(\bar n \cdot P)^2}{Q^2} \frac{w^{\nu}  \varepsilon^{\perp \mu \alpha}\Delta^{\perp}_{\alpha} }{2\eta \bar n \cdot P}.
\end{align}
The only operator that transforms at $\lambda^1$ accuracy is $\f O_{\bar \chi \p \chi}^{[\Gamma]}$. Indeed, we have
\begin{align}
\bar \chi_c(-s \bar n/2) \frac{\Gamma}{2} i\overleftrightarrow{\p}_{\perp}^{\mu} \chi_c(s\bar n/2) &= \bar \chi_c(-s \bar n/2) \frac{\Gamma}{2} i\overleftrightarrow{\p}^{\perp\mu} \chi_c(s\bar n/2) - i\frac{\Delta^{\perp \mu}}{\bar n \cdot \Delta} \frac{d}{ds} \bar \chi_c(-s \bar n/2) \frac{\Gamma}{2} \chi_c(s\bar n/2).
\end{align}
Taking matrix elements and Fourier transforming gives the 
\begin{align}
F_{\bar \chi \p \chi \perp}^{[\Gamma]\mu} = F_{\bar \chi \p \chi}^{[\Gamma]\perp\mu} - \frac{x \Delta^{\perp \mu}}{\bar n \cdot \Delta} F_{\bar \chi \chi}^{[\Gamma]}. 
\end{align}
To compare with the result from \cite{Kivel:2000cn}, we also need to consider the term proportional to $\Delta^{\perp}$ from the equations of motion relation that were dropped in eqs. \eqref{eq: eliminate K} and \eqref{eq: eliminate wK}. Ignoring the quark-gluon-quark contributions, since they cancel anyways, we have
\begin{align}
K^{\perp \mu} &= x F^{\perp \mu} - i \varepsilon^{\perp \mu \nu} \eta \w F^{\perp}_{\nu} - \frac{i}{2} \varepsilon^{\perp \mu \nu} \frac{ \Delta^{\perp}_{\nu} }{\bar n \cdot P} \bar n \cdot \w F,
\\
\w K^{\perp \mu} &= x \w F^{\perp \mu} - i \varepsilon^{\perp \mu \nu} \eta F^{\perp}_{\nu} - \frac{i}{2} \varepsilon^{\perp \mu \nu} \frac{ \Delta^{\perp}_{\nu} }{\bar n \cdot P} \bar n \cdot F. 
\end{align}
Taking into account all these contributions, we find that the shift $\delta T = T - T|_{(n ,\bar n ) \rightarrow (w, \bar w)}$ due to changing the coordinates is
\begin{align}
\delta T^{\mu \nu(0)} = - \intR dx\, \Big ( g^{\perp \mu \alpha} C_- \bar n \cdot F + i \varepsilon^{\perp \mu \alpha}  C_+ \bar n \cdot \w F \Big ) \frac{ \Delta^{\perp}_{\alpha} P^{\nu}}{P \cdot q},
\end{align}
in agreement with \cite{Kivel:2000cn}.

For completeness, we also give the results for the RPI transformation to the Compton frame. It has the same form as in eq. \eqref{eq: n to w} with
\begin{align}
\delta_1^{\mu} =  \frac{2\Delta^{\perp \mu}}{\bar n \cdot \Delta} ,\qquad \delta_2^{\mu} = - \frac{2\Delta^{\perp \mu} \, \bar n \cdot (q+q') }{\bar n \cdot \Delta \, n \cdot (q+q')}.
\end{align}
The result is
\begin{align}
\delta T^{\mu \nu (0)} = - \intR dx\, \Delta^{\perp}_{\alpha}\Bigg \{  C_- \bar n \cdot F \Big ( \frac{g^{\perp \mu \alpha} P^{\nu} - g^{\perp \nu \alpha} P^{\mu}}{2P \cdot q} \Big ) +  C_+ \bar n \cdot \w F \Big (  \frac{ i \epsilon^{\perp \mu \alpha} P^{\nu} + i \epsilon^{\perp \nu \alpha} P^{\mu}}{2P \cdot q} \Big )   \Bigg \},
\end{align}
in agreement with \cite{Belitsky:2005qn}.

\section{Conclusions and Outlook}

We have given an in-depth analysis of the factorization of DVCS at NLP using SCET. We conclude that, at least for DDVCS, where both photons are far off-shell, the amplitude factorizes \textit{to all orders} in terms of twist-3 GPDs in exactly the form that was hypothesized in the literature, as given in eq. \eqref{eq: factorization theorem}. The corresponding SCET operators were given in eqs. \eqref{eq: LP operators} and \eqref{eq: NLP operators}.
In section \ref{sec: calc}, we performed the corresponding tree-level matching in the SCET formalism, and in section \ref{sec: old approach}, we compared and found agreement of the coefficient functions with the known results in the literature. In section \ref{sec: RPI}, the RPI has been used to convert the results into various frames.

The most interesting issue appears when one photon is on-shell, as in DVCS. In section \ref{sec: factorization}, we identified an endpoint-like contribution of parametrical size $\Lambda_{\rm QCD}/Q$, which does factorize but not in terms of a GPD. We have given a precise all-order formula for this contribution in eq. \eqref{eq: Tfig5}. We have also argued that this result is on its face merely technical, since ultrasoft modes are unphysical for exclusive processes. In order to fully understand DVCS at NLP this contribution, which is fundamentally represented by the reduced graph in figure \ref{fig: allorderreduced 2 and 3}(b), must be phenomenologically assessed. This can for example be done using light-cone sum rules. We mention also that due to a quark having small virtuality in the endpoint region we obtain sensitivity to the strange quark mass. These interesting issues are left for future work.

In section \eqref{sec: nontarget regions} we have also proven the already anticipated result to all orders: The standard ``collinear'' NLP contribution involving twist-3 GPDs appears only for the longitudinally polarized virtual photon, in which case it actually constitutes the leading $\sim \lambda^1$ power contribution. It is therefore cleanly separated from the (ultra)soft endpoint-like contribution, which appears only for the transversely polarized virtual photon. Recall that this is only true in the BMP frame. For frames where $\Delta_{\perp} \neq 0$ the LP contribution ``mixes'' into the collinear NLP contribution through RPI transformations. This has been discussed in section \ref{sec: RPI}. 


Although there are many open questions regarding factorization revolving around processes involving GPDs, what SCET can bring to this field of study, particularly beyond the leading power, has not been investigated. This work provides a first step in this direction by applying the framework of the most prominent GPD process, DVCS. 


Moreover, SCET provides a \textit{systematic} approach to the power expansion, which can be straightforwardly generalized beyond the leading order in $\alpha_s$. Going to the next-to-leading order will not only be conceptually interesting, but it will also be relevant for DVCS phenomenology. At this accuracy, one can expect that the genuine twist-3 operators enter, particularly the pure gluon operators in eq. \eqref{eq: NLP operators}.

\section*{Acknowledgements}
We thank Vladimir Braun, Martin Beneke, Yoshitaka Hatta for useful comments.
The U.S. Department of Energy supported this work through Contract No. DE- SC0012704.
J.S. was also supported by Laboratory Directed Research and Development (LDRD) funds from Brookhaven Science Associates.

\appendix

\section{SCET Lagrangian at next-to-leading power}
\label{sec: SCETI L}

The ${\rm SCET_I}$ Lagrangian at $\mathcal{O}(\lambda^1)$ accuracy in position space formalism \cite{Beneke:2002ni} reads
\begin{align} \notag
\f L_{\rm SCET_I} &= \f L_c^{(0)}(\xi_c, A_c,A_{us}) + \f L_{\bar c}^{(0)}(\xi_{\bar c}, A_{\bar c},A_{us}) + \f L_{us}^{(0)}(q_{us}, A_{us}) 
\\
&\quad + \f L_{cus}^{(1)}(\xi_c, q_{us}, A_c, A_{us}) + \f L_{\bar c u s}^{(1)}(\xi_{\bar c}, q_{us}, A_{\bar c}, A_{us}) + \f O(\lambda^2/d^4x),
\end{align}
where \footnote{Greek indices run from $0$ to $4$ as usual, while roman indices, in this case $j$, run from $1$ to $2$.}
\begin{align}
\frac{\lambda^0}{d^4x} \sim \f L_c^{(0)} &= \bar \xi_c \Big ( i n \cdot D_c + g n \cdot A_{us} + i \s D_{c \perp} \frac{1}{i\bar n\cdot D_c} i \s D_{c \perp} \Big ) \frac{\sbar n}{2} \xi_c - \frac{1}{2} \text{tr}( F_c^{\mu \nu} F_{c \mu \nu} )  \notag
\\
\frac{\lambda^0}{d^4x} \sim \f L_{us}^{(0)} &=  \bar q_{us} i \s D_{us} q_{us} - \frac{1}{2} \text{tr}(F_{us}^{\mu \nu} F_{us\mu \nu}) \notag
\\
\frac{\lambda^1}{d^4x} \sim \f L_{cus}^{(1)} &=\bar \xi_c  x_{\perp}^{\mu} n^{\nu} W_c  g F_{s\mu \nu}^{[n]}  W_c^{\dagger}  \frac{\sbar n}{2} \xi_c  + \bar q_{us}^{[n]} W_c^{\dagger} i \s D_{c \perp} \xi_c - \bar \xi_c i \overleftarrow{\s D}_{c \perp} W_c q_{us}^{[n]}
\\ 
&\quad + \text{tr}\Big ( \bar n^{\mu} F_{c\mu j} W_c i \Big [ x_{\perp}^{\rho} n^{\sigma} F_{us\rho \sigma}^{[n]} , W_c^{\dagger} (i D_c^j W_c ) \Big ] W_c^{\dagger}  \Big ) - \text{tr} \Big ( \bar n_{\mu} F_c^{\mu j} W_c n^{\rho} F_{us\rho j}^{[n]} W_c^{\dagger} \Big ). \notag
\end{align}
The Lagrangians $\f L_{\bar c}^{(0)}$ and $\f L_{\bar c us}^{(1)}$ are the same with $n \leftrightarrow \bar n$ and $c \leftrightarrow \bar c$. In the following, the corresponding definitions with $n \leftrightarrow \bar n$ and $c \leftrightarrow \bar c$ can be obtained by this replacement.

A superscript $[n]$ means that the ultrasoft fields are multipole expanded with respect to the $n$-collinear direction
\begin{align}
\phi_{us}^{[n]}(x) = \phi_{us}\Big ( \bar n \cdot x \frac{n}{2} \Big ).
\end{align}
The covariant derivatives are defined as
\begin{align}
D_c^{\mu} = \p^{\mu} - i g A_c^{\mu}, \qquad D_{us}^{\mu} = \p^{\mu} - ig A_{us}^{\mu}.
\end{align}
The collinear and ultrasoft Wilson lines are defined by
\begin{align}
W_c &= P \exp \Big ( ig \int_{-\infty}^0 ds\, \bar n \cdot A_c(x+ s \bar n ) \Big ),
\\
Y_n &= P \exp \Big ( ig \int_{-\infty}^0 ds\, n \cdot A_{us}(x+ sn) \Big ).
\end{align}
The ultrasoft field strength tensor is just the usual field strength tensor expressed in terms of corresponding ultrasoft covariant derivative
\begin{align}
F_{us}^{\mu \nu} = \frac{i}{g} [D_{us}^{\mu} , D_{us}^{\nu} ]
\end{align}
and the collinear field strength tensor is defined as
\begin{align}
F_c^{\mu \nu} = \frac{i}{g} \left [ D_c^{\mu} - ig \frac{\bar n^{\mu}}{2} n \cdot A_{us} , D_c^{\nu} - ig \frac{\bar n^{\nu}}{2} n \cdot A_{us}  \right ].
\end{align}

\bibliographystyle{JHEP}

\bibliography{bibl}%

\end{document}